\documentclass{aa}  

\usepackage{graphicx}
\usepackage{color}
\usepackage{natbib}
\usepackage{savefnmark}
\bibpunct{(}{)}{;}{a}{}{,}
\usepackage{txfonts}
\begin{document} 

\def\sun{\hbox{$\odot$}}
\def\degr{\hbox{$^\circ$}}
\def\arcmin{\hbox{$^\prime$}}
\def\arcsec{\hbox{$^{\prime\prime}$}}

   \title{Molecular tendrils feeding star formation in the Eye of the Medusa}

   \subtitle{The Medusa merger in high resolution $^{\rm 12}$CO\,2$-$1 maps}

   \author{S. K\"onig
          \inst{1}
          \and
          S. Aalto
	   \inst{2}
          \and
          L. Lindroos
	   \inst{2}
          \and
          S. Muller
	   \inst{2}
	  \and
          J.~S. Gallagher III
	   \inst{3}
	  \and
          R.~J. Beswick
	   \inst{4}
	  \and
          G. Petitpas
	   \inst{5}
	  \and
          E. J\"utte
	   \inst{6}
          }

    \offprints{S. K\"onig}

   \institute{Institut de Radioastronomie Millim\'etrique, 300 rue de la Piscine, Domaine Universitaire, F-38406 Saint 
  	     Martin d'H\`eres, France\\
              \email{koenig@iram.fr}
         \and
             Chalmers University of Technology, Department of Earth and Space Sciences, Onsala Space Observatory, 43992 
	     Onsala, Sweden
         \and
             Department of Astronomy, University of Wisconsin, 475 N. Charter Street, Madison, WI, 53706, USA
         \and
             University of Manchester, Jodrell Bank Centre for Astrophysics, Oxford Road, Manchester, M13 9PL, UK
         \and
             Harvard-Smithsonian Center for Astrophysics, 60 Garden Street, Cambridge, MA, 02138, USA
	 \and 
	     Astronomisches Institut Ruhr-Universit\"at Bochum, Universit\"atsstra\ss e 150, 44780 Bochum, Germany
             }

   \date{Received ; accepted }

 
\abstract
 {Studying molecular gas properties in merging galaxies gives us important clues to the onset and evolution of interaction-triggered 
 starbursts. NGC~4194 (the Medusa merger) is particularly interesting to study since its FIR-to-CO luminosity ratio rivals 
 that of ultraluminous galaxies (ULIRGs), despite its lower luminosity compared to ULIRGs, which indicates a high star formation 
 efficiency (SFE) that is relative to even most spirals and ULIRGs.
 We study the molecular medium at an angular resolution of 0.65\arcsec\,$\times$\,0.52\arcsec\ ($\sim$120\,$\times$\,98~pc) 
 through our observations of $^{\rm 12}$CO\,2$-$1 emission using the Submillimeter Array (SMA). We compare our $^{\rm 12}$CO\,2$-$1 maps 
 with optical \textit{Hubble Space Telescope} and high angular resolution radio continuum images to study the relationship between 
 molecular gas and the other components of the starburst region. 
 The molecular gas is tracing the complicated dust lane structure of NGC~4194 with the brightest emission being located in an off-nuclear 
 ring-like structure with $\sim$320~pc radius, ``the Eye of the Medusa''. The bulk CO emission of the ring is found south of the 
 kinematical center of NGC~4194. The northern tip of the ring is associated with the galaxy nucleus, where the radio continuum 
 has its peak. Large velocity widths associated with the radio nucleus support the notion of NGC~4194 hosting an AGN. A prominent, 
 secondary emission maximum in the radio continuum is located inside the molecular ring. This suggests that the morphology of the 
 ring is partially influenced by massive supernova explosions. From the combined evidence, we propose that the Eye of the Medusa 
 contains a shell of swept up material where we identify a number of giant molecular associations (GMAs). We propose that the Eye may 
 be the site of an efficient starburst of 5-7~M$_{\sun}$\,yr$^{\rm -1}$, but it would still constitute only a fraction of the 
 30-50~M$_{\sun}$\,yr$^{\rm -1}$ star formation rate of the Medusa.
 Furthermore, we find that $\sim$50\% of the molecular mass of NGC~4194 is found in extended filamentary-like structures tracing the 
 minor and major axis dust lanes. We suggest that molecular gas is transported along these lanes providing the central starburst region 
 with fuel. Interestingly, a comparison with locations of ``super star clusters'' (SSCs) reveal that the molecular gas and the SSCs are 
 not co-spatial.}


\titlerunning{High-resolution CO\,2$-$1 in NGC~4194}

   \maketitle
%

\section{Introduction}
The focus on merger studies often lies on major-major (equal mass spirals) mergers and their evolution, although minor (unequal-mass) 
mergers constitute the major phase of interactions in the local Universe and at higher redshifts. Understanding how the gas is feeding 
starburst and AGN activities in these objects is therefore paramount in understanding the overall evolution of the Universe. In a 
numerical simulation study of minor- or intermediate mergers, \citet{bour05} found that the gas brought in by the disturbing companion 
galaxy is generally found at large radii in the merger remnant. The gas returns to the system from tidal tails and often forms 
rings - polar, inclined or equatorial – that will appear as dust lanes when seen edge-on \citep[e.g.,][]{com88,shl89}.\\
\indent
Evidence has been presented for an increased fraction of star formation happening in clusters within starbursts that are compared to that 
in quiescent galaxies, which potentially are linked to the interstellar medium (ISM) structure changes, while the starburst process takes 
place \citep{ken12}. Massive, young star clusters (super star clusters (SSCs), 10$^{\rm 5}$ - 10$^{\rm 8}$~M$_{\sun}$) are observed to 
form in on-going starbursts, such as those associated with interactions and (minor and major) mergers and those in smaller numbers in 
other types of galaxies with elevated star formation rates \citep[SFR; e.g.,][]{lar02,deG03,mor09}.\\
\begin{figure*}[t]
  \begin{minipage}[hbt]{0.4925\textwidth}
  \centering
    \includegraphics[width=0.75\textwidth,angle=-90]{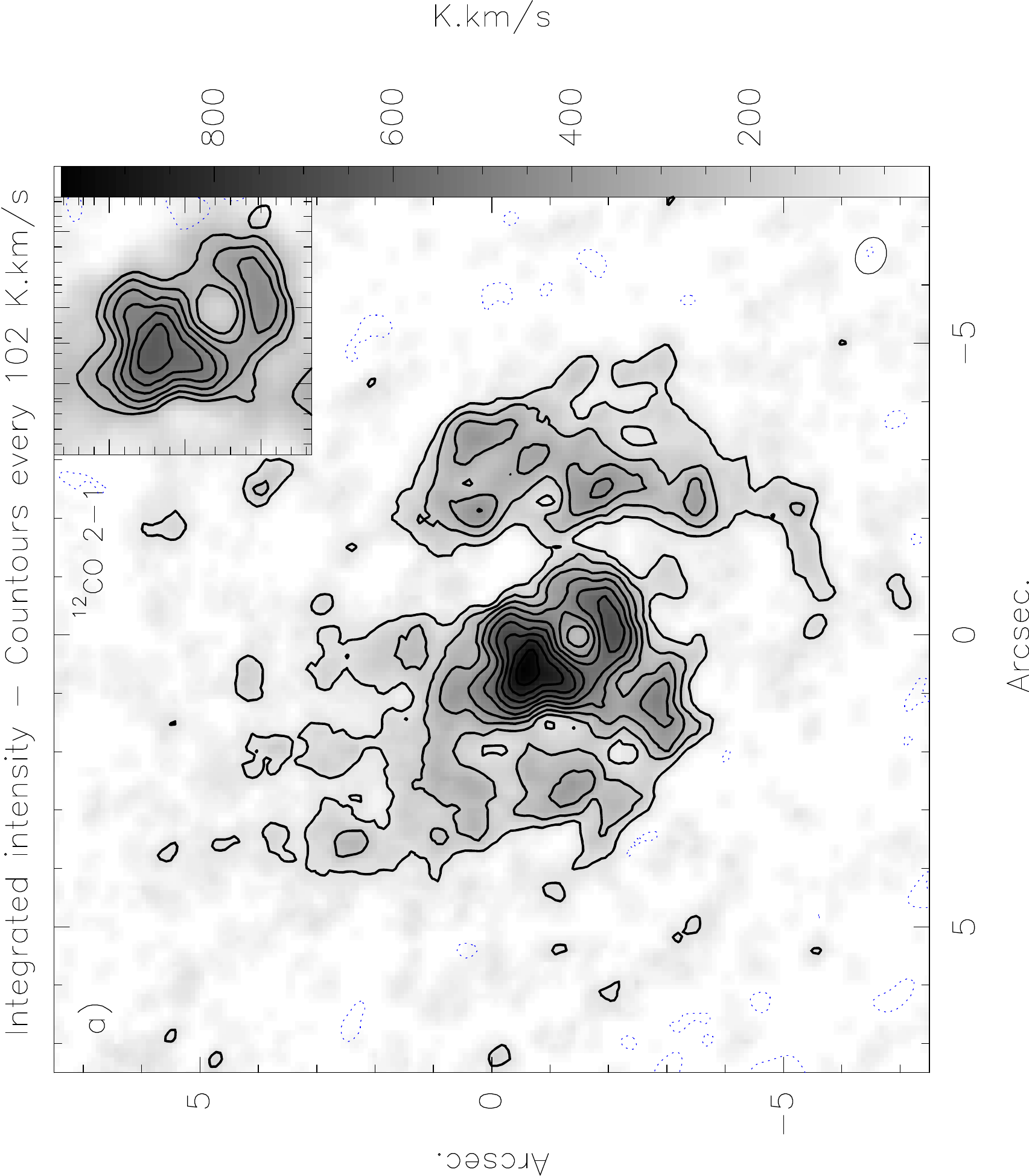}
    \includegraphics[width=0.75\textwidth,angle=-90]{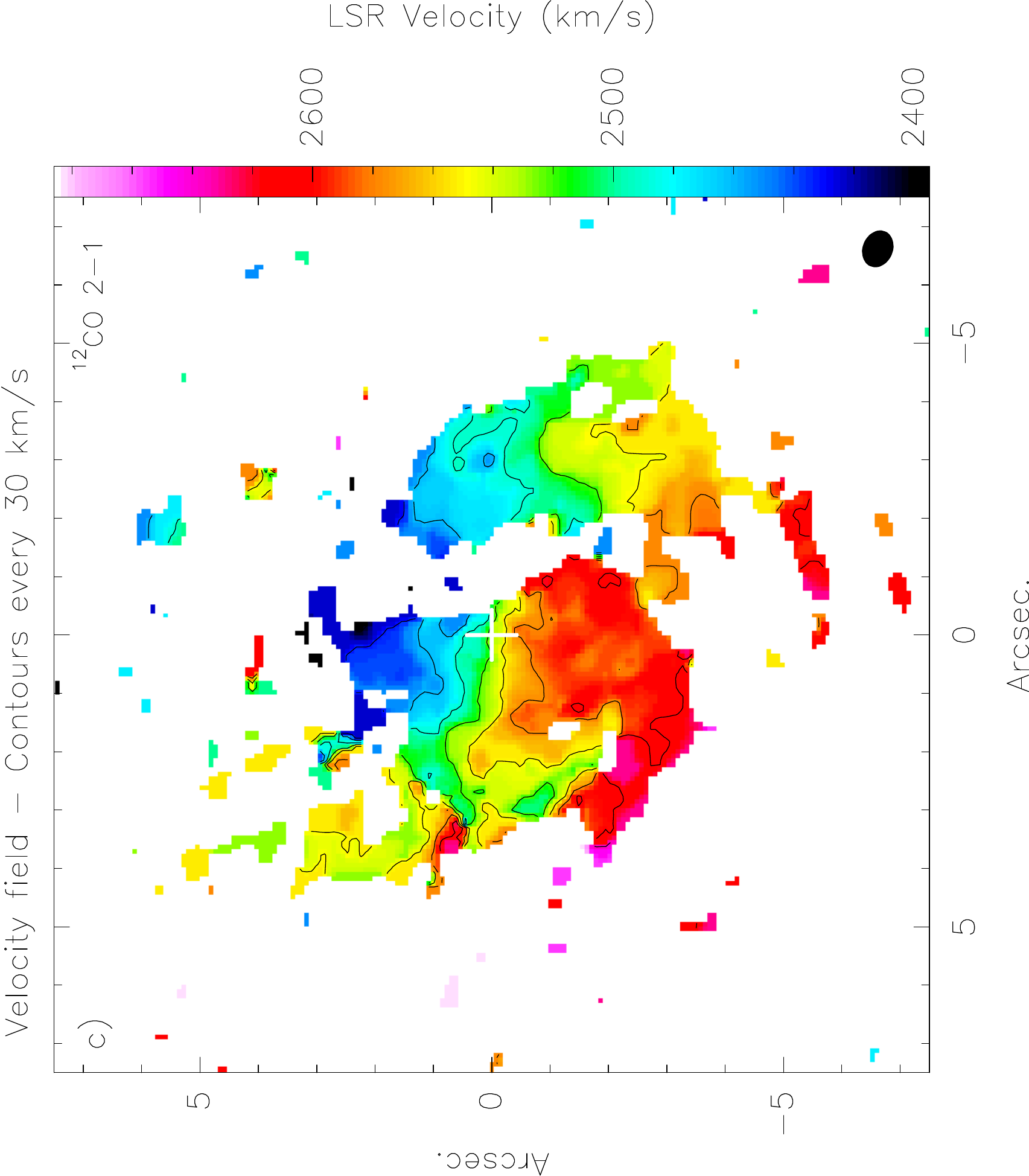}
  \end{minipage}
  \begin{minipage}[hbt]{0.4925\textwidth}
   \hspace{-0.55cm}
  \centering
    \includegraphics[width=0.75\textwidth,angle=-90]{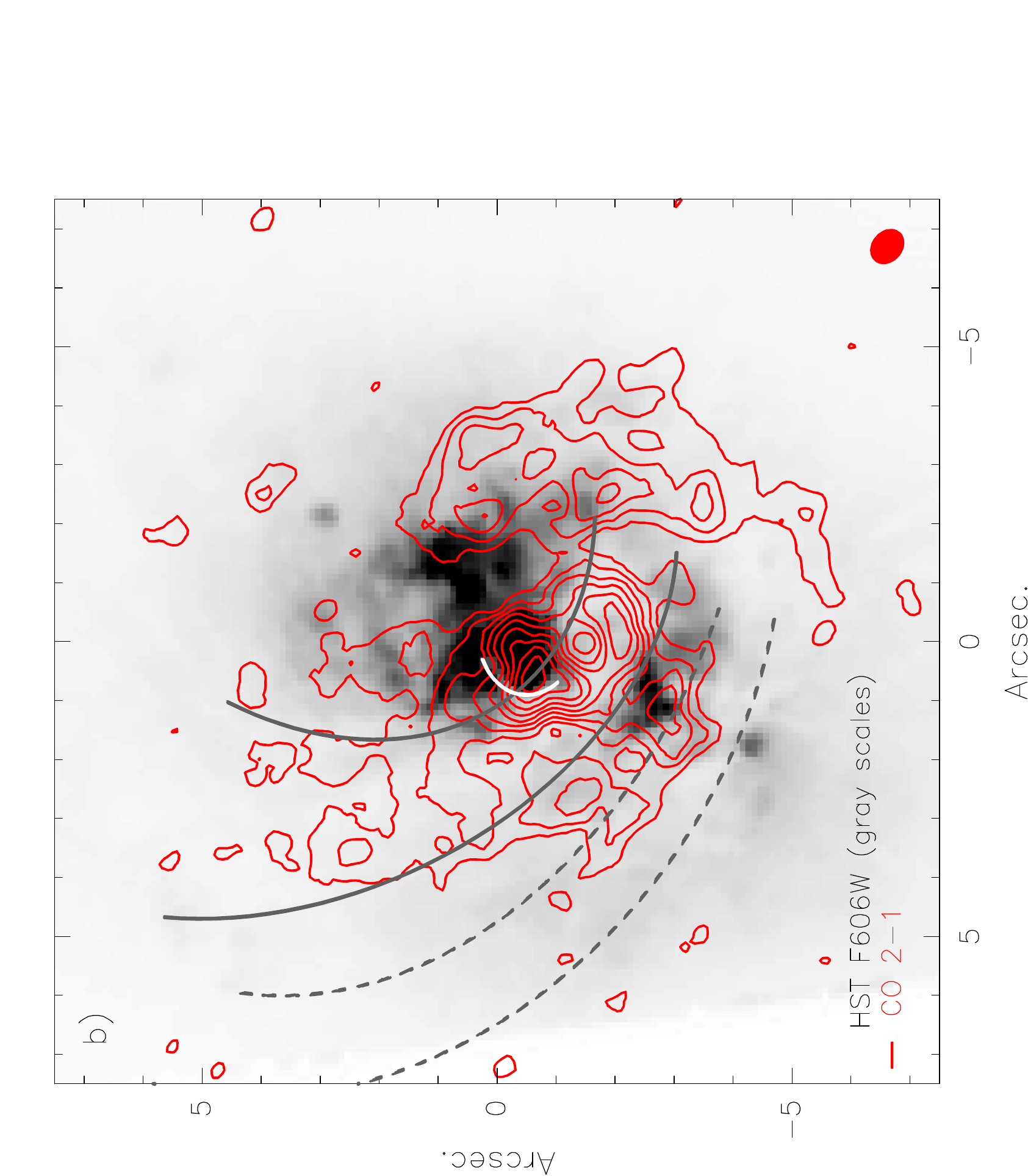}
    \includegraphics[width=0.75\textwidth,angle=-90]{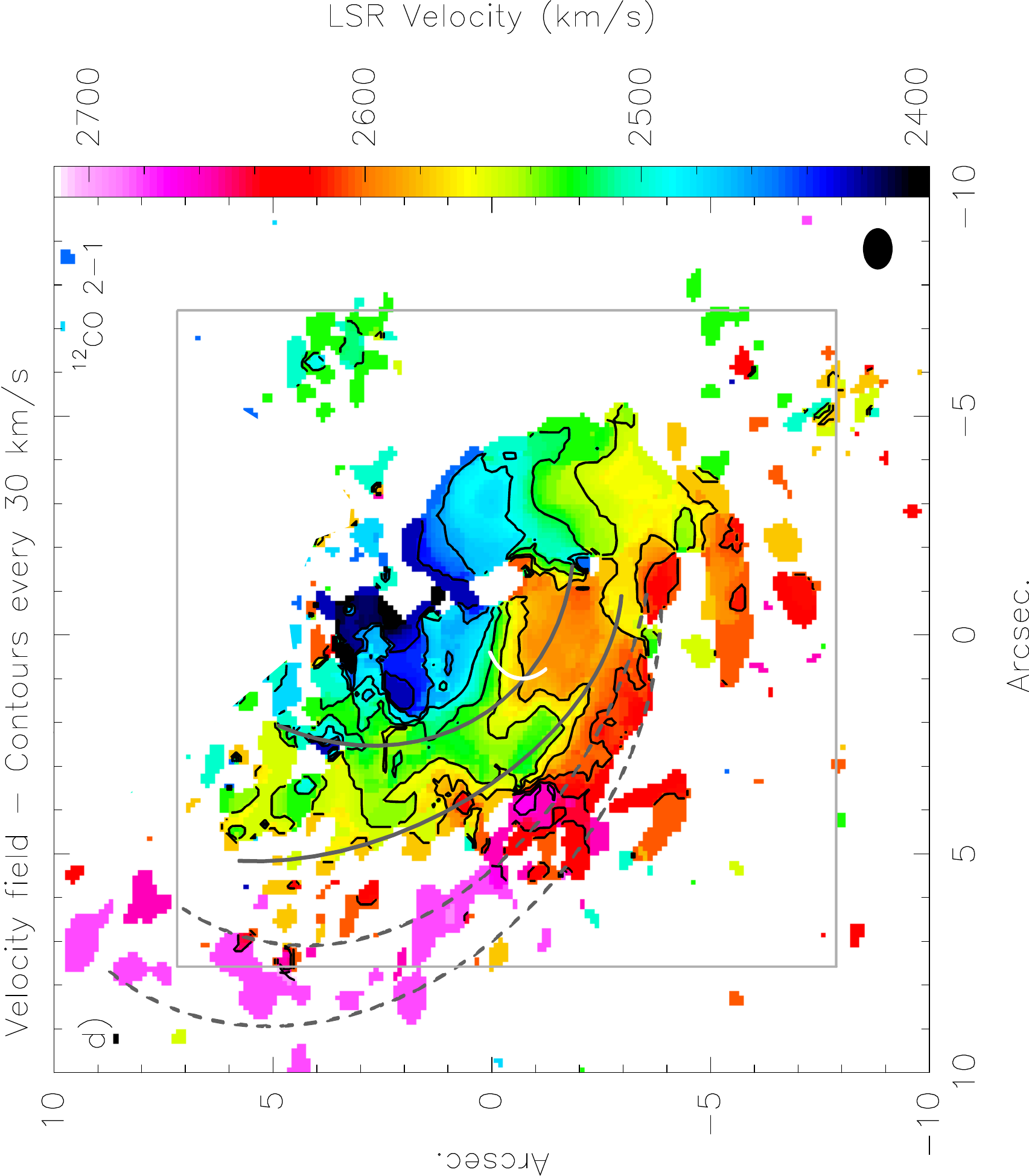}
  \end{minipage}
  \caption{\footnotesize Integrated intensity map (\textit{upper left}) and velocity field (\textit{bottom left}) of the high resolution 
   CO\,2$-$1 emission (0.65\arcsec\,$\times$\,0.52\arcsec), velocity field of the low resolution CO\,2$-$1 emission (\textit{bottom 
   right}, 0.95\arcsec\,$\times$\,0.68\arcsec) in NGC~4194, and an overlay of the high resolution $^{\rm 12}$CO\,2$-$1 emission contours 
   on top of an HST WFPC2 F606W filter image (\textit{top right}). The noise level in the integrated intensity map is 
   1.3~Jy\,km\,s$^{\rm -1}$. Contours for the integrated intensity map start at 2$\sigma$ in steps of 2$\sigma$; the first negative 
   contour at -2$\sigma$ is shown as a dotted contour. The velocity range plotted in the velocity field is between 2400~km\,s$^{\rm -1}$ 
   and 2700~km\,s$^{\rm -1}$; contours range from 2400~km\,s$^{\rm -1}$ to 2640~km\,s$^{\rm -1}$. The gray, gray-dashed, and white curves 
   in the HST overlay and lower-resolution velocity field represent the locations of the most important dust lanes. The solid line box in 
   the lower-resolution velocity field represents the field-of-view shown in the higher-resolution velocity field. North is up, east to 
   the left. The position of the 1.4~GHz continuum peak is marked by a white cross \citep{bes05}.}
  \label{fig:co2-1}
\end{figure*}
\indent
A nearby example of a surprisingly efficient starburst is the inner 2~kpc of the \object{Medusa merger} (\object{NGC~4194}). With a 
luminosity of L$_{\rm FIR}$\,=\,8.5\,$\times$\,10$^{\rm 10}$~L$_{\sun}$ (at D\,=\,39~Mpc, 1\arcsec\,=\,189~pc), this E+S minor merger is 
an order of magnitude fainter than well known ULIRGs, such as \object{Arp~220} \citep{aalto00,man08}. Despite the moderate FIR 
luminosity, NGC~4194 has a L$_{\rm FIR}$/L$_{\rm CO}$ ratio similar to those typical for ULIRGs, which suggests that its star formation 
efficiency (SFE) rivals that of the compact ULIRGs. Despite this, its CO/HCN\,1$-$0 luminosity ratio is $\sim$25, indicating that the 
fraction of dense gas is significantly lower despite the similar SFE to ULIRGs. The picture becomes even more interesting when one 
considers that most of the ongoing star formation in \object{NGC~4194} is not traced by the FIR or radio emission: The H$\alpha$ SFR is 
$\sim$40~M$_{\sun}$\,yr$^{-1}$ \citep{han06}, while the FIR estimated SFR is 6-7~M$_{\sun}$\,yr$^{-1}$. The spatial correlation between 
the molecular gas distribution and the 1.4~GHz continuum is also poor \citep[unlike the case for most nearby galaxies,][]{aalto00}. In 
\object{NGC~4194}, a large fraction of the star formation is going on in super star clusters with a kpc-scale distribution, which is 
separated from the molecular gas distribution. No obvious correlation between these young SSCs (5-15~Myr) and the CO can be found. 
The CO emission also traces the two prominent dust lanes that cross the central region and extends into the northern tidal tail. The 
majority of the CO \citep[$\sim $70$\%$, $^{\rm 12}$CO,][]{aalto01,lin11} is found in the central 2~kpc of the galaxy with 15$\%$ of this 
gas \citep[$^{\rm 13}$CO,][]{aalto10}, which resides in a compact region 1.5\arcsec\ south of the radio nucleus. The knots where 
the optically traced star formation is going on only occasionally correlate with the radio continuum or the molecular distribution.\\
\indent
Throughout the paper, we are concerned with pure rotational transitions in CO between upper state $J'$\,=\,2 and lower state $J''$\,=\,1 
that are labeled 2$-$1.\\
\indent
In this paper, we present a study of the molecular gas close to the AGN in the \object{Medusa merger}. In Sect. \ref{sec:obs}, we 
describe the observations and data reduction; Sect. \ref{sec:results} reports on the results of the observations. The discussion follows 
in Sects. \ref{sec:discussion_1} and \ref{sec:discussion_2}, and conclusions are drawn in Sect. \ref{sec:summary}.


\begin{figure*}[ht]
  \begin{minipage}[hbt]{0.4925\textwidth}
  \centering
    \includegraphics[width=0.75\textwidth,angle=-90]{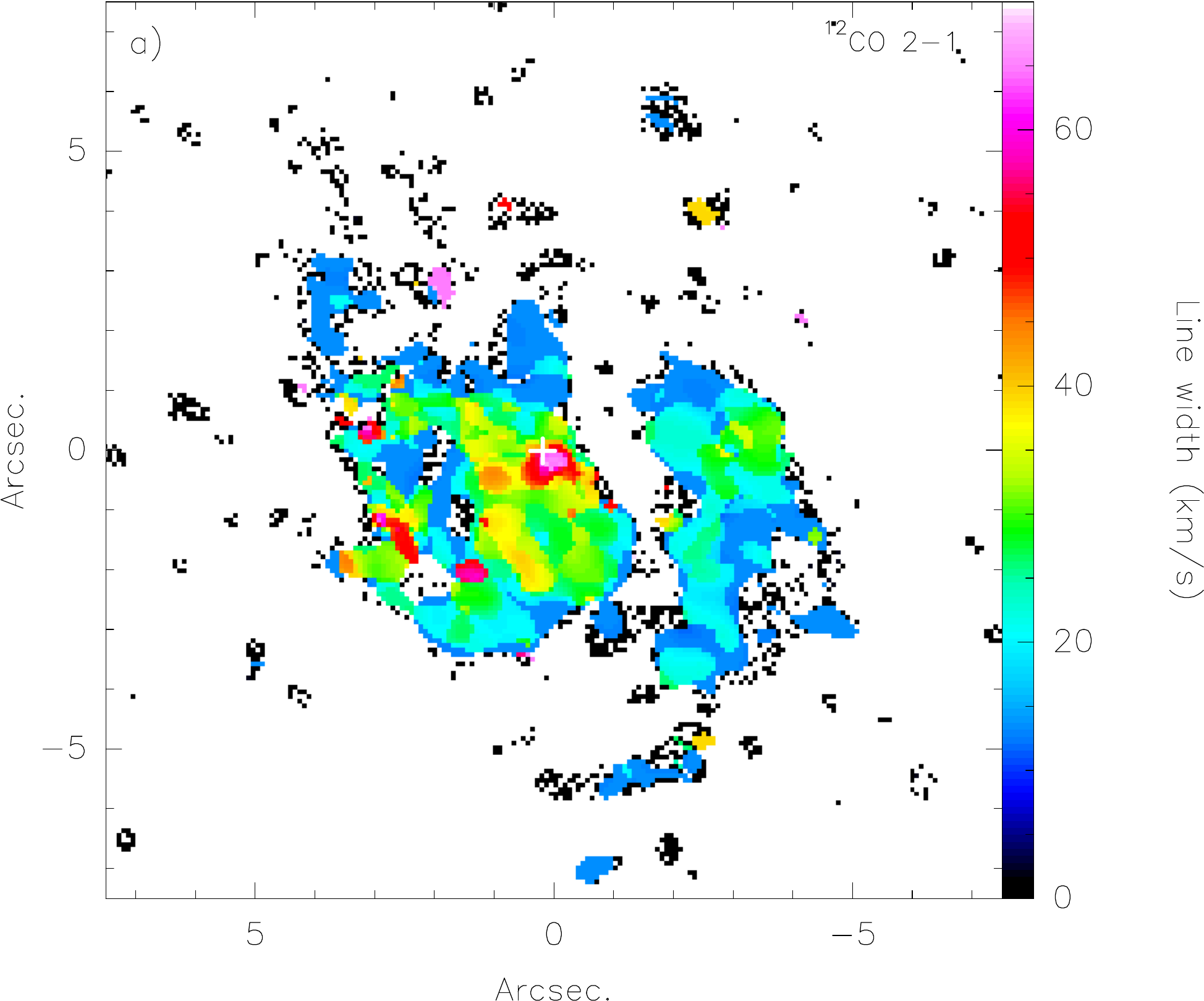}
  \end{minipage}
  \begin{minipage}[hbt]{0.4925\textwidth}
  \centering
    \includegraphics[width=0.75\textwidth,angle=-90]{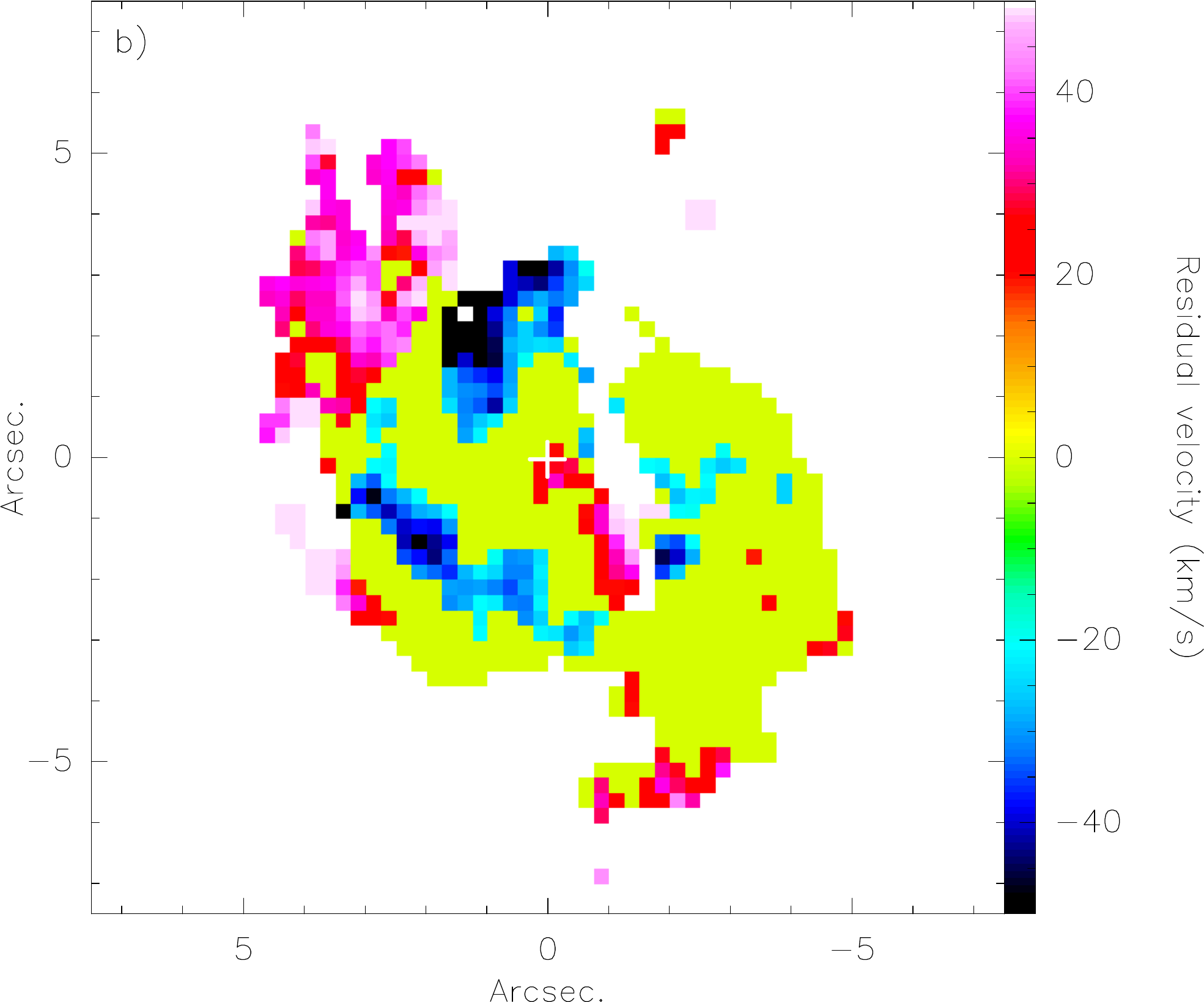}
  \end{minipage}
  \caption{\footnotesize Velocity dispersion map of the high-resolution CO\,2$-$1 emission (\textit{left}). Map showing the 
   residuals in the velocity field after the subtraction of the rotation curve from the CO velocity field (\textit{right}). Residuals in 
   the range $\pm$20~km\,s$^{\rm -1}$, which corresponds to the noise limit in the residuals map, are plotted in light green; regions 
   with values larger than that are depicted in red (positive values) and blue (negative values). The white crosses mark the peak 
   position of the radio continuum from \citet{bes05}.}
  \label{fig:rot_curve_residuals}
\end{figure*}

\section{Observations} \label{sec:obs}

%
\object{NGC~4194} was observed with the SubMillimeter Array (SMA) on February 21, 2010 in the very-extended configuration and on April 
8 in the compact configuration, which provides the highest angular resolution CO observations of NGC~4194 to date. For analysis 
purposes, we shifted the phase center to the position of the 1.4~GHz radio continuum peak at $\alpha$=12:14:09.660 and 
$\delta$=+54:31:35.85 \citep[J2000,][]{bes05}. The heterodyne receivers were tuned to the redshifted frequency of the 
$^{\rm 12}$CO\,2$-$1 transition at 228.63~GHz in the upper sideband, while the $^{\rm 13}$CO\,2$-$1 transition was observed in the lower 
sideband. With baseline lengths between 38~m and 509~m, these SMA imaging data are sensitive to scales smaller than 16.5\arcsec. 
The correlator was configured to provide a spectral resolution of 0.8125~MHz (corresponding to a velocity resolution of 
$\sim$1.1~km\,s$^{\rm -1}$). The bright quasars \object{J0854+201} and \object{3C454.3} were used as bandpass calibrators; \object{Vesta} 
and \object{Ganymede} were observed as primary flux calibrators, and we regularly observed the close-by quasars \object{J1153+72}, 
\object{J0927+390}, and \object{J0721+713} for complex gain calibration.\\
\indent
After calibration within the dedicated MIR/IDL SMA reduction package, both visibility sets were converted into FITS format and imported 
in the GILDAS/MAPPING\footnote{http://www.iram.fr/IRAMFR/GILDAS} and AIPS packages for further imaging.\\
\indent
For the $^{\rm 12}$CO\,2$-$1 data, sets of visibilities from the compact and very extended configuration observations were combined and 
deconvolved using the Clark method \citep{cla80} with robust weighting. This results in a synthesized beam size of 
0.65\arcsec\,$\times$\,0.52\arcsec\ with a position angle (PA) of 67\degr. We smoothed the data to a velocity resolution of 
$\sim$25~km\,s$^{\rm -1}$, which yields a 1$\sigma$ rms noise level per channel of $\sim$6.8~mJy\,beam$^{\rm -1}$. To look at the most 
compact component in the CO distribution, we used robust weighting, putting additional weight on the longest baselines, resulting 
in a 0.43\arcsec\,$\times$\,0.38\arcsec\ beam with a position angle of 52\degr. This provides a very high resolution map in which we look 
for giant molecular associations (GMAs).\\
\indent
To compare our results with high resolution images of the radio continuum structure of NGC~4194, we obtained 1.4~GHz radio 
continuum data at the VLA (J\"utte et al., in prep.) and combined them with published data from \citet{bes05}. These combined data have 
an angular resolution of 0.50\arcsec\,$\times$\,0.52\arcsec\ corresponding to a linear scale of $\sim$95pc\,$\times$\,98pc.\\


\section{Results} \label{sec:results}

\subsection{$^{\rm 12}$CO\,2$-$1} \label{subsec:12CO}

\begin{figure*}[ht]
  \begin{minipage}[hbt]{0.4925\textwidth}
  \centering
    \includegraphics[width=0.775\textwidth,angle=-90]{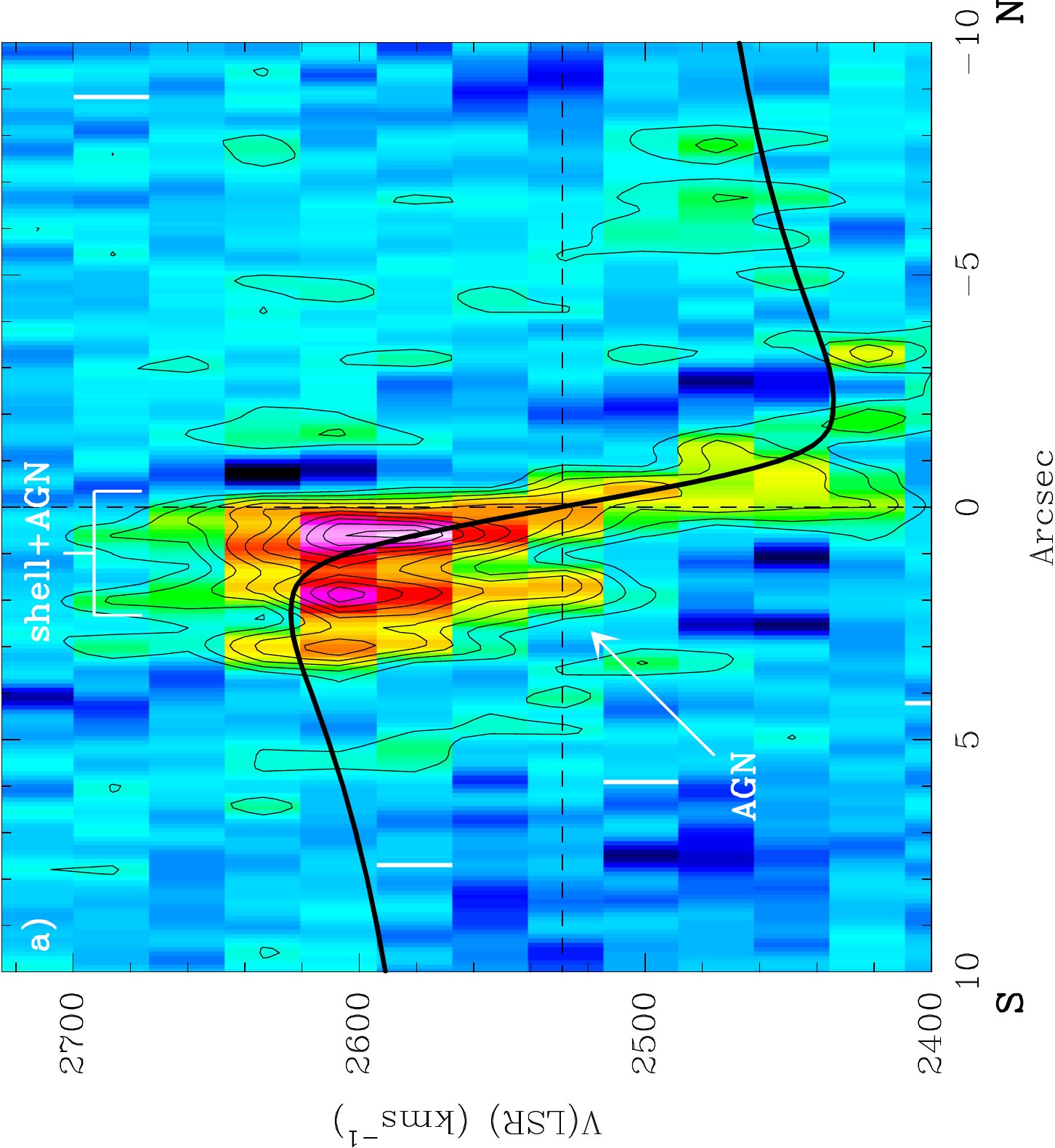}
  \end{minipage}
  \begin{minipage}[hbt]{0.4925\textwidth}
  \centering
    \includegraphics[width=0.775\textwidth,angle=-90]{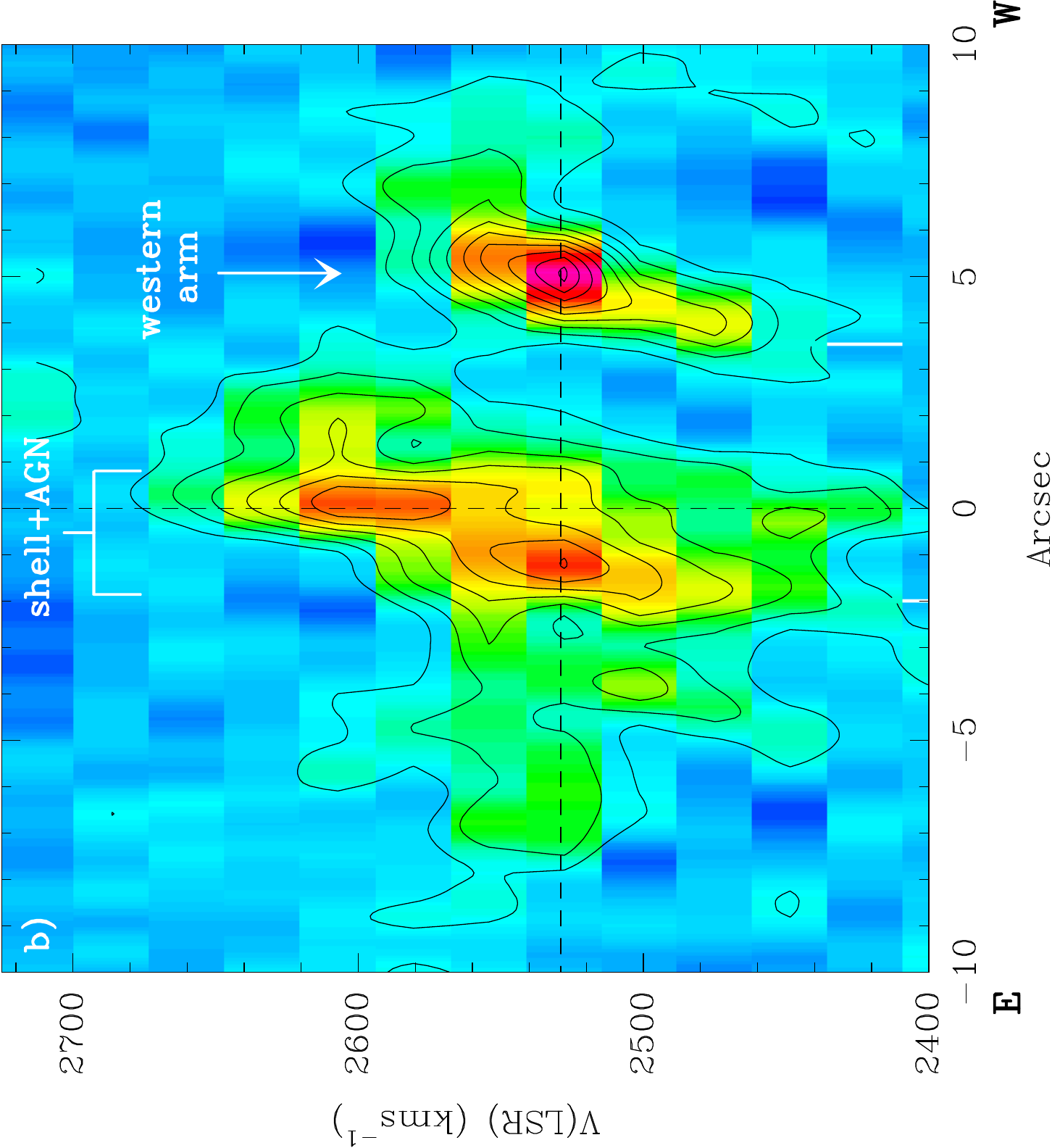}
  \end{minipage}
  \caption{\footnotesize CO\,2$-$1 position–velocity diagrams along the major (north-south, \textit{left}) and the minor (east-west, 
   \textit{right}) axes of the NGC~4194 obtained from the high-resolution data cube within a slit width of 1\arcsec. The contour levels 
   start at 1.5$\sigma$ in steps of 1.5$\sigma$. The colors range from 0 to 0.1~mJy/beam. The result of the fitted Brandt rotation curve, 
   using a fixed inclination value of 40\degr, is shown on the left.}
  \label{fig:co2-1_pv}
\end{figure*}

\subsubsection{Integrated intensity} \label{subsec:integr_intensity}

The integrated intensity map (Fig.\,\ref{fig:co2-1}a) shows an asymmetric CO\,2$-$1 distribution in the east-west direction. The eastern 
part contains the peak of the CO emission at about 1-1.5\arcsec\ southeast of the radio continuum peak, which is located in the northern 
part of a ring-like structure or possibly tightly wound spiral arms, with a maximum extent of $\sim$4\arcsec\ that has no CO emission 
peak at its center (Fig.\,\ref{fig:co2-1}a).\\ 
\indent
The total CO\,2$-$1 flux within the 2$\sigma$ contours (see Fig.\,\ref{fig:co2-1}a) amounts to 411.0\,$\pm$\,4.1~Jy\,km\,s$^{\rm -1}$. A 
comparison of the total CO\,2$-$1 flux within the inner 13\arcsec\ of NGC~4194 obtained from single-dish observations \citet{cas92} shows 
that we recover approximately 50\% ($\sim$642~Jy\,km\,s$^{\rm -1}$) of the flux with our SMA observations. Our high-resolution data are 
sensitive to structures smaller than 16.5\arcsec\ (see Sect.\,\ref{sec:obs}), hence 50\% of the flux seems to be associated with 
structures larger than this size. Approximately one third of the total flux is located in the east of the ring-like structure 
(128.3\,$\pm$\,4.5~Jy\,km\,s$^{\rm -1}$) and one fourth (S$_{\rm \nu}$\,=\,103.7\,$\pm$\,2.7~Jy\,km\,s$^{\rm -1}$) in the elongated 
western part of the CO\,2$-$1 distribution.\\
\indent
\citet{cas92} show the integrated intensity of CO\,2$-$1 to be 80\% of that of CO\,1$-$0. Thus, we can scale our CO\,2$-$1 flux by this 
amount to obtain an estimated CO\,1$-$0 flux for our observations. Assuming a CO-to-H$_{\rm 2}$ conversion factor of 
$X_{\rm CO}$\,=\,2.0\,$\times$\,10$^{\rm 20}$~cm$^{\rm -2}$\,(K\,km\,s$^{\rm -1})$$^{\rm -1}$ \citep{nar12,san13}\footnote{We will study 
the influence of the distribution of molecular gas on $X_{\rm CO}$ through different molecular transitions in paper II (K\"onig et al. in 
prep.).}, this then translates to a mass of $\sim$1.6\,$\times$\,10$^{\rm 9}$~M$_{\sun}$ for the overall CO\,2$-$1 molecular gas 
distribution. The molecular gas in the ring sums up to a mass of $\sim$4.1\,$\times$\,10$^{\rm 8}$~M$_{\sun}$.

\subsubsection{Kinematics of the $^{\rm 12}$CO\,2$-$1 gas} \label{subsec:kinematics}

Figs.\,\ref{fig:co2-1}c, d, and \ref{fig:rot_curve_residuals} show the velocity field and the velocity dispersion of the CO\,2$-$1 
emission in NGC~4194. \textit{The velocity fields} (Figs.\,\ref{fig:co2-1}c, d) show a rather regular pattern for the main body of the 
molecular emission, which is typical for solid body rotation, with a position angle of -20\degr\ in which the velocities range from 
2400~km\,s$^{\rm -1}$ to 2700~km\,s$^{\rm -1}$. The contours of the velocity field in the western component of the molecular gas 
emission, which we call the ``western arm'', deviate from the typical solid-body rotation pattern. These deviations could be indicators 
for streaming motions, as found in other galaxies, such as in M~51 and NGC~5248 \citep{aalto99,TvdL13}. \citet{aalto00} have shown that 
molecular gas is associated with the dust lanes that cross NGC~4194 in front of the main body. The velocities of this gas emission are 
almost constant across large distances along the way to the center. This behavior changes just before the gaseous material turns into the 
plane of the merger (see, e.g., the most eastern CO associated with the eastern dust lane,as seen in Fig.\,\ref{fig:co2-1}c). This seems 
to be true as well for the dust lane located further west and closer to the center of NGC~4194.\\
\indent
\textit{The velocity dispersion} distribution in the Medusa (Fig.\,\ref{fig:rot_curve_residuals}a) shows the most prominent peak at the 
position of the radio continuum peak \citep{bes05} with a line dispersion roughly twice the size than in the rest of the CO distribution. 
The line dispersion at this peak rises to about 70~km\,s$^{\rm -1}$ compared to the surrounding material where the dispersion lies 
between 15~km\,s$^{\rm -1}$ and 35-40~km\,s$^{\rm -1}$.\\
\indent
\textit{The position-velocity diagrams} along the major (north to south, left panel) and the minor (east to west, right panel) axes are 
shown in Fig.\,\ref{fig:co2-1_pv}. Both pv diagrams were obtained by averaging over slits with a width of 1\arcsec\ with one positioned 
along the axis of solid-body rotation and the other perpendicular to that direction. Both slits do include emission from parts of the 
molecular ring-like structure. In the major axis pv diagram (Fig.\,\ref{fig:co2-1_pv}a), the pattern of the rather smooth solid-body 
rotation, especially when considering the merger history, already seen in the velocity field (Figs.\,\ref{fig:co2-1}c, d) has been 
reproduced. The two most distinctive peaks in this distribution represent the structure of the ring-like gas component at the center of 
NGC~4194. The minor axis pv diagram (Fig.\,\ref{fig:co2-1_pv}b) shows two distinct components in the distribution: one molecular emission 
peak with a steep velocity gradient across a few arcseconds in the west and two distribution peaks close to the nucleus. The feature with 
the steep velocity gradient originates from gas in the same location that already showed an exceptional behavior in the velocity field 
(Figs.\,\ref{fig:co2-1}c, d) in the western arm. This might hint toward the presence of streaming motions in the western arm. The second 
component close to the nucleus represents the gas in the ring-like structure by showing a difference in velocity for the different 
locations in this structure.\\
\indent
\textit{The rotation curve:} We fitted a Brandt curve \citep[Eq.\,\ref{eq:brandt_curve},][]{bra60} to the velocity field of the molecular 
emission using a fixed inclination value of 40\degr\ \citep[from the optical isophotes,][]{aalto00}. The result is shown in 
Fig.\,\ref{fig:co2-1_pv}a with the pv diagrams for the major and minor axes. The fitted rotation curve starts to flatten (fall) at a 
radius R$_{\rm max}$ of 2.3\arcsec, which indicates a deprojected rotational velocity v$_{\rm rot}$ of 148~km\,s$^{\rm -1}$. The 
dynamical mass inferred from these values is M$_{\rm dyn}$\,$\sim$2.2\,$\times$\,10$^{\rm 9}$~M$_{\sun}$. A plot of the residuals after 
subtracting the fit from the velocity field is presented in Fig.\,\ref{fig:rot_curve_residuals}b. There, the residuals in the western 
arm again illustrate the presence of typical features of streaming motions in this location. We cannot draw clear conclusions whether or 
not some type of streaming motion is also going on in the eastern part of the molecular emission as well due to the likely overlap of 
different structures and processes taking place there.
\begin{equation}
 \centering
  v_{\rm rot}\left( R \right) = \frac{v_{\rm max} \left( R/R_{\rm max} \right)}{\left[ \frac{1}{3}+\frac{2}{3} \left( R/R_{\rm max} \right)^{\rm n} \right]^{\rm \frac{3}{2n}}} \ .
  \label{eq:brandt_curve}
\end{equation}

\subsection{The Eye of the Medusa} \label{subsec:eye_of_medusa}

\begin{figure}[ht]
  \centering
    \includegraphics[width=0.4\textwidth,angle=-90]{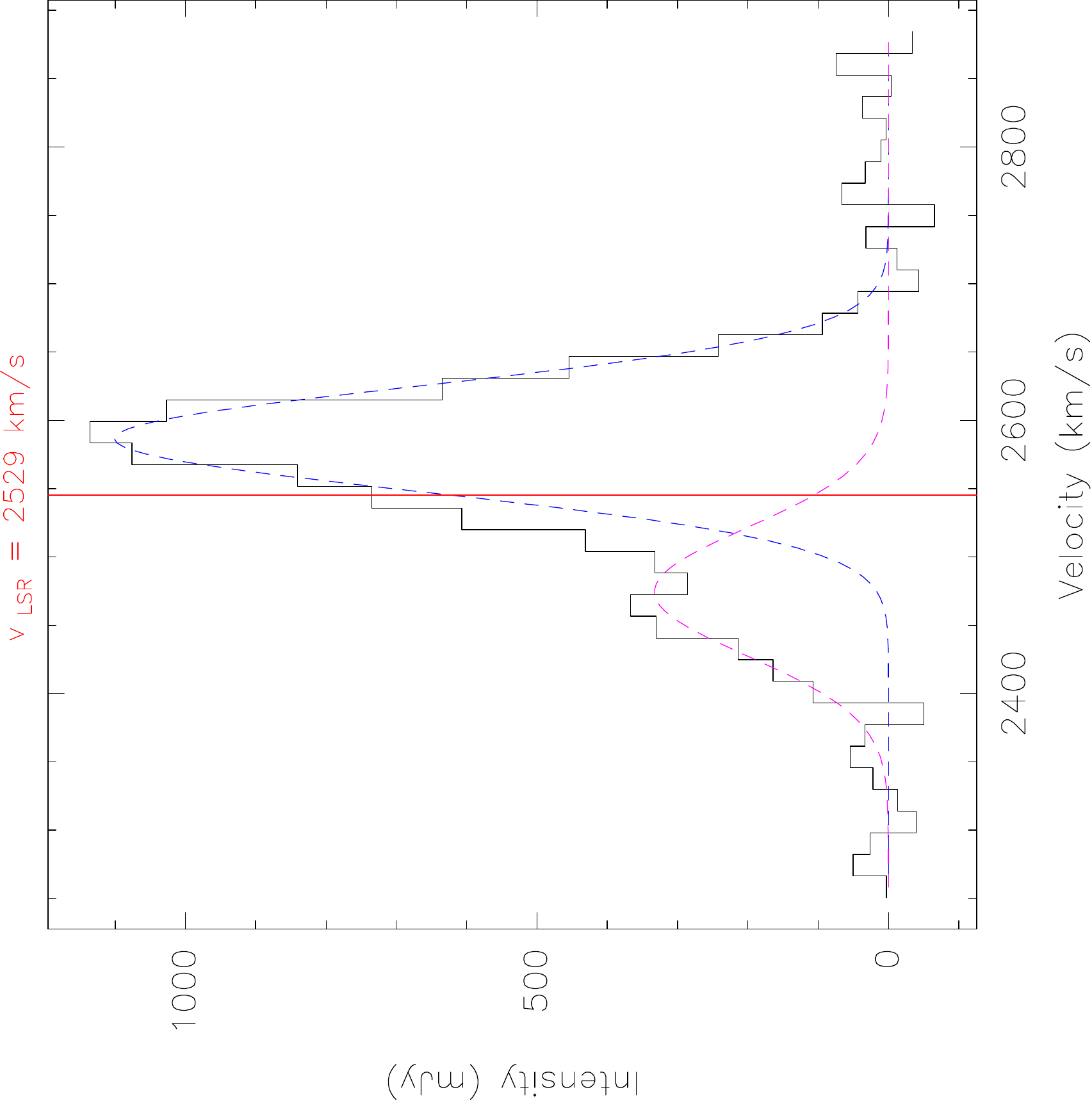}
  \caption{\footnotesize Spectrum of the central high density gas region (Eye of the Medusa) containing the AGN and the shell. 
   A Gaussian fit shows that the spectrum is well reproduced by two velocity components (magenta and blue). The velocity resolution is 
   $\sim$15~km\,s$^{\rm -1}$. A closer view at the spectra at different positions in the central region shows that the gas represented by 
   the velocity component centered at 2450~km\,s$^{\rm -1}$ (magenta) is only present close to the nucleus of NGC~4194.}
  \label{fig:co2-1_highres_spectrum}
\end{figure}

Figs.\,\ref{fig:co2-1}a and b show a top-heavy ring-like structure, where the majority of the molecular mass is located in the northern 
part of the structure. This structure dominates the CO\,2$-$1 emission in the eastern part of NGC~4194, which is close to its center. We 
determined the ring-like structure to have a molecular mass of $\sim$4.1\,$\times$\,10$^{\rm 8}$~M$_{\sun}$ (corresponding to one third 
of the total CO\,2$-$1 mass in these data), enclosed in a radius of $\sim$1.7\arcsec\ (equivalent to $\sim$320~pc) with a width of 
1.5\arcsec\ ($\sim$285~pc) in the northern part of the ring and a width of 0.9\arcsec\ ($\sim$170~pc) in the south. The nucleus of 
NGC~4194 is located at the northern tip of the ring, while the bulk emission of the ring is located south of the kinematical center in 
a molecular shell or bubble. In contrast to 1.4~GHz radio continuum observations \citep[Fig.\,\ref{fig:overlays_highres_ring},][]{bes05}, 
where a secondary peak is located at the center of the ring, the CO\,2$-$1 does not show a clear emission peak at its center. This 
structure was named the Eye of the Medusa \citep{lin11}.\\
\noindent
The CO\,2$-$1 emission distribution of the Eye, shows ``hot spots'' of enhanced $^{\rm 12}$CO\,2$-$1 intensity 
(Fig.\,\ref{fig:co2-1_ring}). The mass estimates of some of those clumps place them in the mass range for giant molecular associations 
\citep[GMAs, Table\,\ref{tab:GMA_properties},][]{vog88}. We identified GMA candidates as peaks (peak values larger than three times the 
noise level) in the integrated CO\,2$-$1 map and with a clear detection ($\gtrsim$\,5$\sigma$) in the spectrum. The spectra were obtained 
by averaging over an area of 5\,$\times$\,5 pixels around the peak. This resulted in the identification of 11 GMA candidates. How many of 
these candidates are real is an issue that can only be solved by higher resolution, high-sensitivity observations. The molecular 
properties of the Eye are discussed in Sect.\,\ref{sec:discussion_1}.


\section{Molecular structure of NGC~4194} \label{sec:discussion_1}

\begin{figure*}[ht]
 \centering
  \begin{minipage}[hbt]{0.4925\textwidth}
  \centering
    \includegraphics[width=0.82\textwidth,angle=-90]{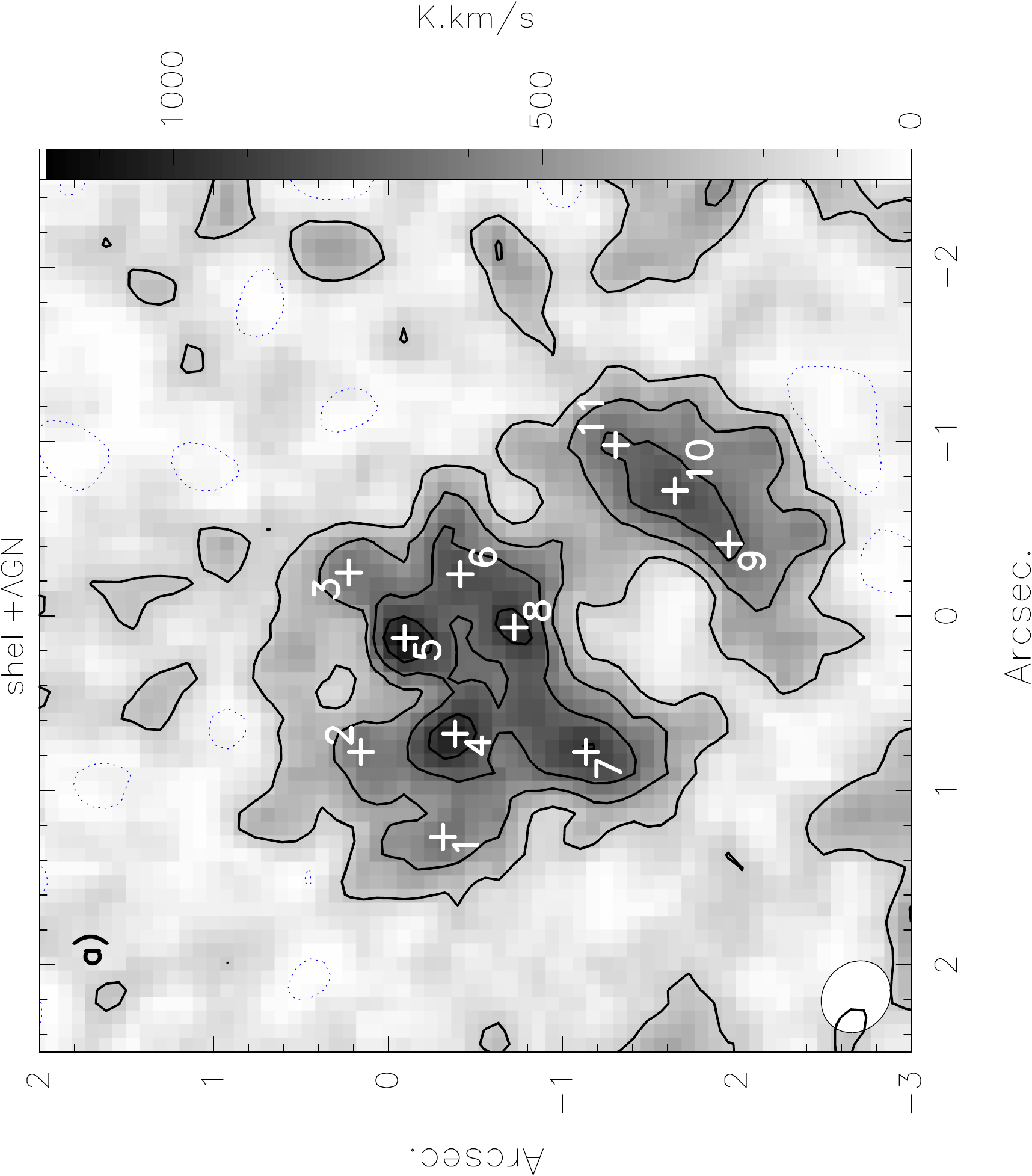}
  \end{minipage}
  \begin{minipage}[hbt]{0.4925\textwidth}
  \centering
    \includegraphics[width=0.82\textwidth,angle=-90]{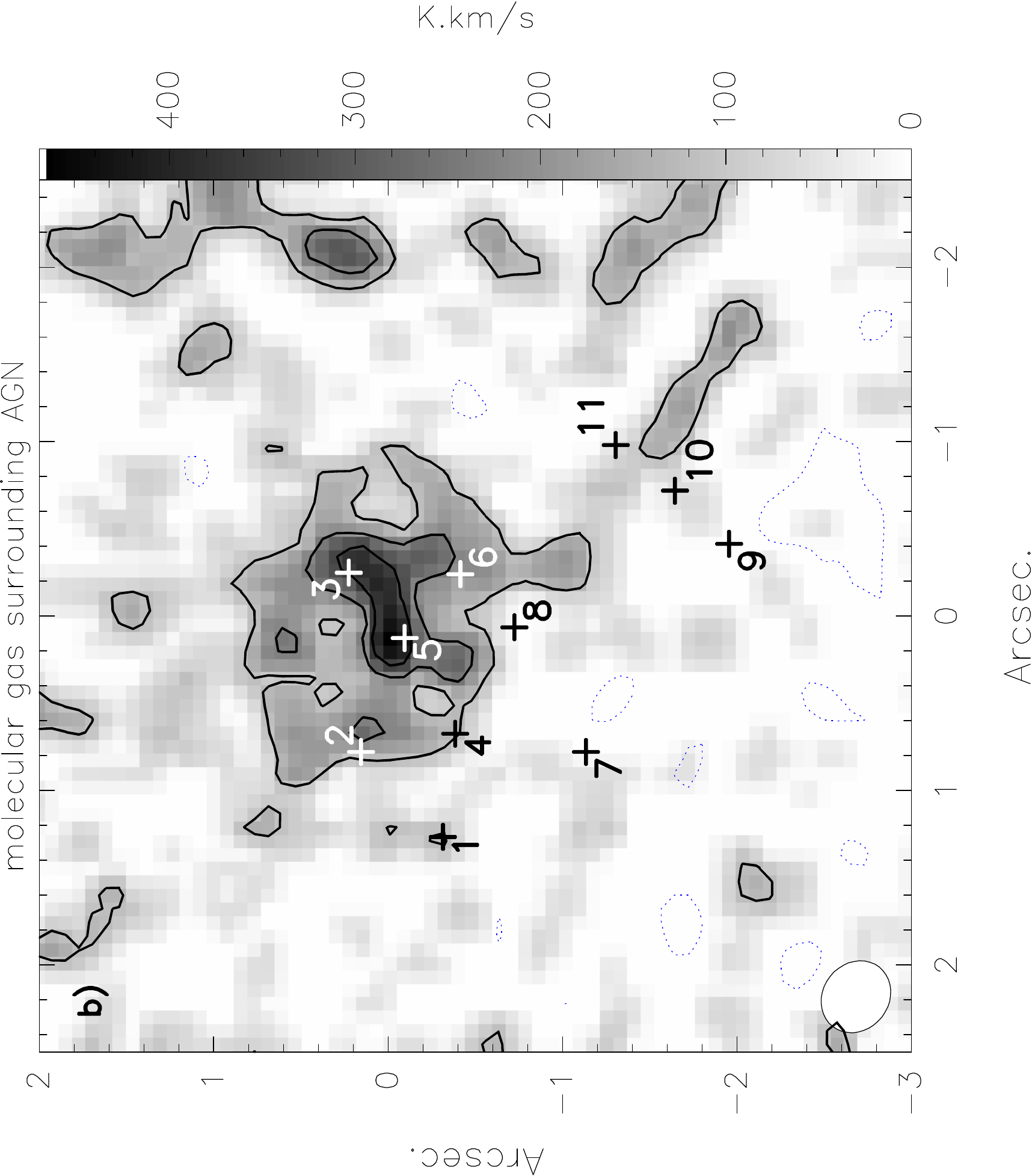}
  \end{minipage}
  \caption{\footnotesize Distribution of the central CO\,2$-$1 emission in high resolution (\textit{left}) and intensity distribution of 
   the gas surrounding the AGN, which is associated only with the second velocity component in the CO\,2$-$1 spectrum (\textit{right}). 
   Crosses and numbers from 1 to 11 mark the positions of the GMAs. GMA\,5 is located at the peak position of the 1.4~GHz radio continuum 
   \citep{bes05}. The beam (0.43\arcsec\,$\times$\,0.38\arcsec) is shown in the lower left corner of the panels. The contours are at 
   every 1.0~Jy\,km\,s$^{\rm -1}$ starting at 1.0~Jy\,km\,s$^{\rm -1}$,; blue dotted contours are the negative counterparts of the lowest 
   positive contour.}
  \label{fig:co2-1_ring}
\end{figure*}

Two major dust lanes to the east of the nucleus, which are visible in the optical (e.g., HST, Fig.\,\ref{fig:co2-1}b), cross the main 
optical body of NGC~4194. The one crossing directly below the galaxy's dynamical center is associated with large parts of the molecular 
gas. The second dust lane further to the east crosses further south of the nucleus. The molecular gas is clearly associated with both 
dust lanes tracing filamentary-like features (see the lower resolution velocity field in Fig.\,\ref{fig:co2-1}d). The CO emission in the 
northeast stretching out toward these dust lanes appears filamentary. The western part of the CO distribution with a north-south 
orientation agrees well with peaks c and d found by \citet{aalto00} in their CO\,1$-$0 data (see Fig.\,\ref{fig:comparison_co1-0_co2-1}). 
There seems to be a gap in the CO distribution between the molecular gas in the east and the west, apart from a small connection between 
the two parts with a width of $\sim$1\arcsec. Within the asymmetric CO\,2$-$1 emission distribution, the ring-like structure, the Eye of 
the Medusa, shows hot spots of enhanced $^{\rm 12}$CO\,2$-$1 intensity (Fig.\,\ref{fig:co2-1_ring}). A more detailed discussion of these 
features can be found in Sect.\,\ref{subsubsec:eye}.\\
\indent
Connected to the CO in the Eye seems to be a gas component that shows the highest velocity dispersion at the 1.4~GHz radio continuum 
peak found by \citet{bes05}. A weak AGN component might be present at this position in the center of NGC~4194. This is discussed in more 
detail in Sect.\,\ref{subsec:agn}.\\
\indent
Observations of different star formation tracers have been reported covering the larger scale structures associated with the CO\,1$-$0 
observations and the dust lanes down to the region surrounding the nucleus of NGC~4194. \cite{wei04} and \citet{han06} found a number of 
stellar clusters in the UV and visible light (VIS, see, e.g., Figs.\,\ref{fig:hst+clusters+co21}, 
\ref{fig:comparison_co2-1_rings}). The majority of these clusters does not seem to be closely associated with the molecular gas. Indeed, 
most of these clusters are positioned in a void of molecular gas between the eastern and western parts in the CO\,2$-$1 emission. For 
further details, see Sect.\,\ref{subsubsec:sf_molecular_gas}.\\
\indent
In Sect.\,\ref{subsubsec:comparison_minor_mergers}, we compare the properties of NGC~4194 to other minor mergers and, in 
Sect.\,\ref{subsubsec:comparison_shells}, we compare our results for NGC~4194 with galaxies that harbor molecular shells and blowouts.

\subsection{The AGN} \label{subsec:agn}

\begin{figure}[t]
  \centering
    \includegraphics[width=0.4\textwidth,angle=-90]{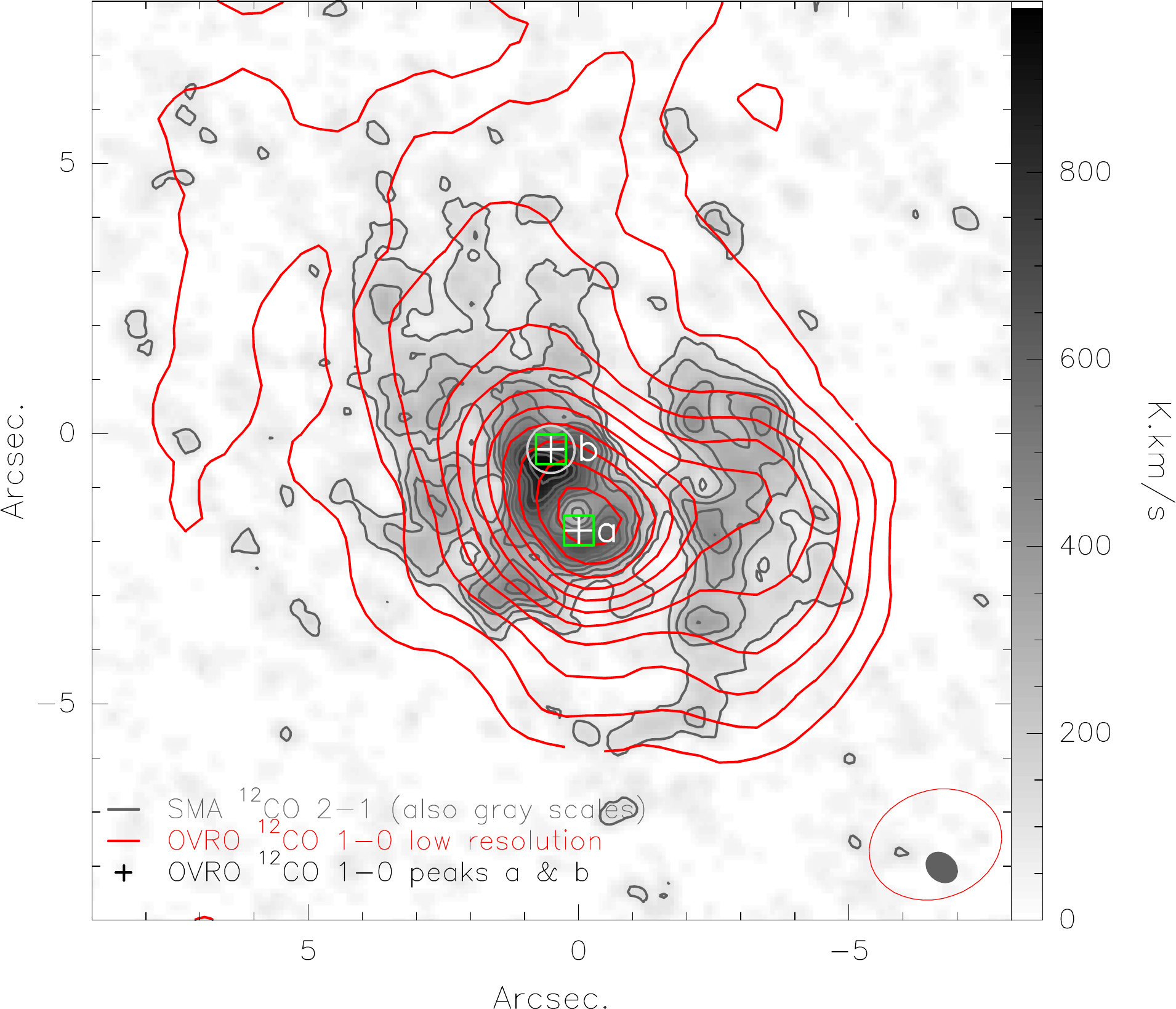}
  \caption{\footnotesize Comparison of the SMA $^{\rm 12}$CO\,2$-$1 emission (gray scales) with OVRO $^{\rm 12}$CO\,1$-$0 low 
   resolution data \citep{aalto00}. The beam sizes are indicated in the lower right corner. White crosses mark the positions of 
   peaks ``a'' and ``b'' found by \citet{aalto00} in their high resolution data, squares represent the CO peaks located in the dust lane 
   and the circle marks the peak position with a counterpart in H$\alpha$ \citep{arm90}.}
  \label{fig:comparison_co1-0_co2-1}
\end{figure}

Several indicators speak in favor of the presence of an AGN in NGC~4194, even if it might be weak. X-ray and [Ne V] 14.3~$\mu$m line 
observations have shown the possibility for the presence of a small AGN contribution to the energy output of NGC~4194. \citet{kaa08} and 
\citet{leh10} suggest the presence of an AGN in the nucleus based on their 2$-$10 and 2$-$8~keV point source identifications and nuclear 
count rates. Spitzer observations of the [Ne V] 14.3~$\mu$m line yielded detections for \citet{ber09} and \citet{leh10} which leads them 
to conclude the presence of a weak AGN being located in NGC~4194.\\
\indent
Spectra taken of the Eye from the high resolution data (0.43\arcsec\,$\times$\,0.38\arcsec) show a double-peaked velocity 
distribution (Fig.\,\ref{fig:co2-1_highres_spectrum}) with velocity components centered at $\sim$2600~km\,s$^{\rm -1}$ and at 
$\sim$2450~km\,s$^{\rm -1}$. The latter component is only present in the region closest to the AGN (GMAs 3, 5 \& 6). The gas mass 
determined from this velocity component is $\sim$4.4\,$\times$\,10$^{\rm 7}$~M$_{\sun}$. Isolating this velocity component and 
making an integrated intensity map confirms that this gas is only present in the northern part of the central high density gas complex 
(Fig.\,\ref{fig:co2-1_ring}b). The total gas mass surrounding the AGN (both CO velocity components) results to 
$\sim$1.1\,$\times$\,10$^{\rm 8}$~M$_{\sun}$. The dynamical mass enclosed in this region is 
$\sim$5.0\,$\times$\,10$^{\rm 8}$~M$_{\sun}$. With their HI absorption observations, \citet{bes05} were able to put a limit on the 
dynamical mass of the central nuclear region of $\sim$2\,$\times$\,10$^{\rm 9}$~M$_{\sun}$, which is in good agreement with the 
results from this work. This CO\,2$-$1 complex is located at the unresolved center of the larger scale molecular gas reservoir in 
NGC~4194 \citep[CO\,1$-$0, see Fig.\,\ref{fig:comparison_co1-0_co2-1},][]{aalto00}.\\
\indent
Speaking against the presence of an AGN, however, are the findings from \citet{bec14}. Their radio continuum and [\ion{Ne}{ii}] 
observations indicate the compact radio emission sources in NGC~4194 to be dense stellar clusters.

\subsection{The structure of the Eye} \label{subsubsec:eye}

\begin{figure}[t]
  \centering
    \includegraphics[width=0.4\textwidth,angle=-90]{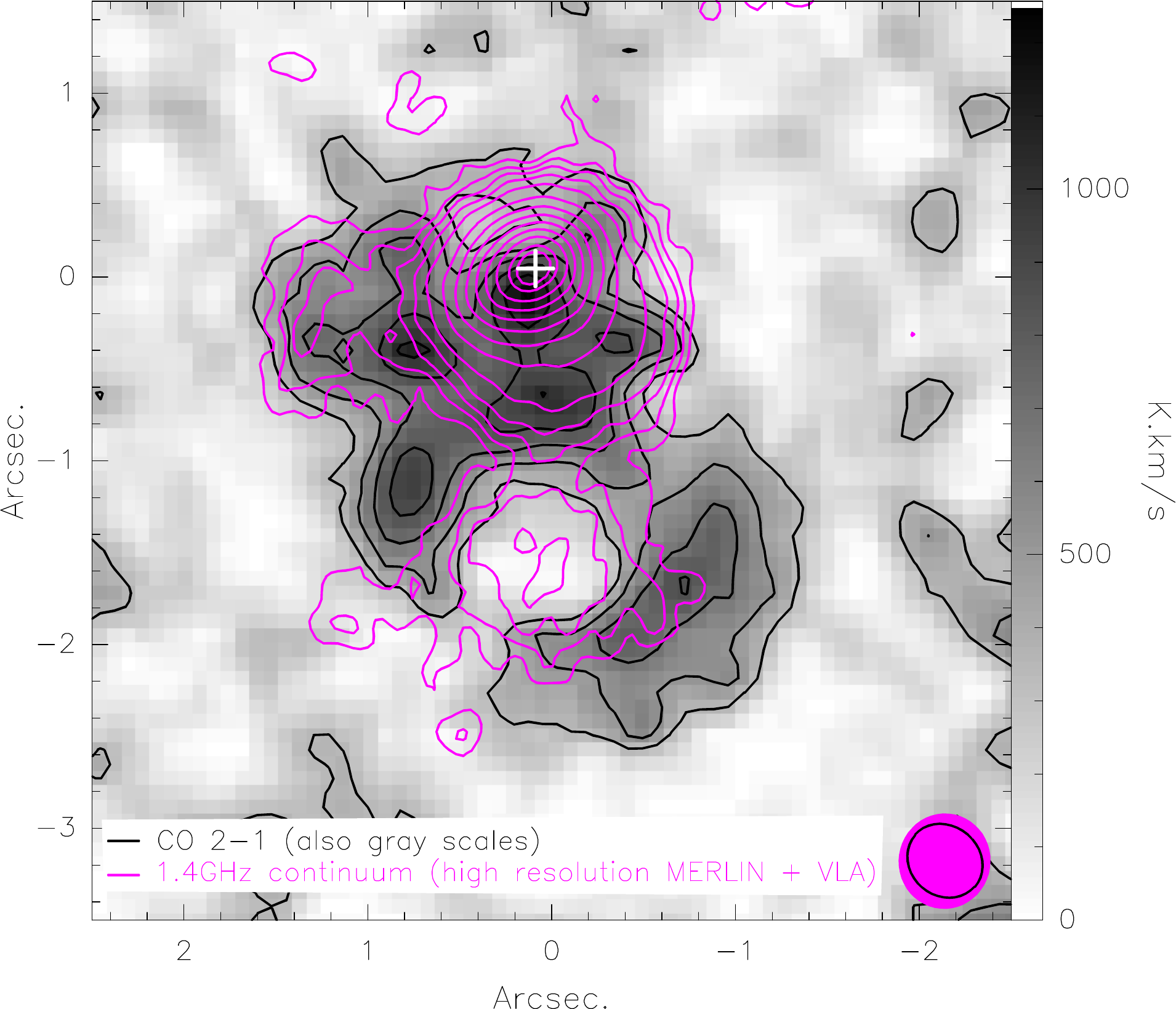}
  \caption{\footnotesize Comparison of the high-resolution CO\,2$-$1 emission with the high-resolution 1.4~GHz emission in the central 
   region of the Medusa. Black contours and gray scale background represent the higher resolution CO\,2$-$1 emission 
   (0.43\arcsec\,$\times$\,0.38\arcsec). The high-resolution combined MERLIN and VLA data are represented by magenta contours. The 
   cross marks the position of the 1.4~GHz radio continuum peak of the MERLIN observations by \citet{bes05}. The beam sizes are depicted, 
   according to the colour schemes, in the lower right corner.}
  \label{fig:overlays_highres_ring}
\end{figure}

The highest surface brightness CO\,2$-$1 emission emerges in a ring-like structure south of the nucleus (Fig.\,\ref{fig:co2-1_ring}b) - 
the Eye of the Medusa. The molecular gas in this complex seems to be connected to the 1.4~GHz radio continuum peak, which possibly marks 
the position of a weak AGN. Most of the gas, however, is associated with the eastern dust lane that comes in closest to the galactic 
center (Fig.\,\ref{fig:co2-1}). The center of the Eye is not located at the dynamical center of NGC~4194, as determined from the 
rotation curve (see Sect.\,\ref{subsec:kinematics}). The location of the possible AGN in the northwestern part of this structure agrees 
with the center coordinates. This implies that this structure might not have been formed from purely dynamical processes (e.g., 
orbit crowding, resonances).\\
\indent
In Figs.\,\ref{fig:co2-1_ring} and \ref{fig:overlays_highres_ring} we show that there is a ``hole'' in the CO emission. Located right at 
the center of that hole is a secondary 1.4~GHz radio continuum peak \citep[Fig.\,\ref{fig:overlays_highres_ring}, see also][]{bes05}. The 
radio continuum very nicely traces the peak of the gas in the northern part of the Eye but it seems to avoid gas in the southern part: 
the radio continuum peaks at the center of the structure where there is no CO, but most of the southern part of the CO in this region is 
not traced by the 1.4~GHz emission as is GMA~7 in the north. The radio continuum associated with GMA~8 most likely spills over from the 
peak at the AGN position.\\
\indent
The fact that the CO depression in the south of the ring-like structure corresponds to a secondary peak in the 1.4~GHz emission and the 
maximum in HI absorption against the continuum \citep{bes05}, suggests that the hole in the CO\,2$-$1 distribution might be a shell or 
bubble in the molecular gas caused by a massive explosion of supernovae (SNe). The funneling of the gas to the very center of NGC~4194 
leeds to a burst of star formation in the central molecular gas. Subsequently, supernovae explode and the energy freed by the SNe going off drove the molecular gas outwards, which is away from the center of the explosion, thereby shaping the central molecular gas reservoir 
to the observed morphology: the expansion of the shell/bubble causes a spherical gas pile-up at the shock fronts, where new star 
formation is triggered and pinpoints the interaction between the accelerated shell/bubble material and the surrounding ISM. The result is 
the Eye of the Medusa, the molecular ring-like structure with a shell or bubble that we observe in the CO\,2$-$1 emission.
\begin{equation}
   E_{\rm 0} = 5.3 \times 10^{\rm -7} n_{\rm 0}^{\rm 1.12} v_{\rm sh}^{\rm 1.4} R^{\rm 3.12} \ .
  \label{eq:chevalier}
\end{equation}
\indent
To derive the energy output necessary for the formation of the shell/bubble with the properties we observe in our CO\,2$-$1 data, we 
applied Chevalier's equation \citep[Eq.\,\ref{eq:chevalier},][]{che74}. Using a hydrogen number density of the surrounding pre-bubble 
medium of $\sim$59~cm$^{\rm -3}$ (value determined from the density and volume of the surrounding gas not yet influenced by the expansion 
of the shell), a radius of 180~pc (the inner radial expansion of the shell material), and an expansion velocity of 55~km\,s$^{\rm -1}$ 
(from the pv diagram), we derive an energy output of 1.4\,$\times$\,10$^{\rm 55}$~ergs, which is the equivalent of $\sim$10\,000 SNe of 
type II. We derived the kinematical age, using $t_{\rm kin}$\,$\approx$\,$\alpha\,R/v$ where $R$ is the radius of the shell or bubble, 
$v$ is the expansion velocity, and $\alpha$ is a parameter to account for nonlinear expansion. We assume $\alpha$ to be $\sim$0.5, a 
value in the middle of the range of possible values \citep{saka06}. Due to the uncertainty in the assumption of $\alpha$, we expect an 
uncertainty in the time estimate of a factor of 2. Hence, the kinematical age for the shell in NGC~4194 is 
$\sim$1.6\,$\times$\,10$^{\rm 6}$~yr, and the supernova rate thus amounts to $\sim$6\,$\times$\,10$^{\rm -3}$~SN\,yr$^{\rm -1}$. A 
comparison to other galaxies hosting shells/bubbles is discussed in Sect.\,\ref{subsubsec:comparison_shells}.\\
\indent
Our data mapped at the highest resolution (beam: 0.43\arcsec\,$\times$\,0.38\arcsec, PA\,=\,52\degr, 1\arcsec\,=\,189~pc) revealed a 
clumpy structure in the Eye component of the CO\,2$-$1 distribution, as in \object{NGC~1097}, \object{NGC~1365}, or \object{NGC~1614}, 
for example. We identified 11 possible GMAs (see Fig.\,\ref{fig:co2-1_ring}); the majority (GMA~1 to GMA~8) are located in the northern 
part of the structure. Masses M(H$_{\rm 2}$) between $\sim$1.0\,$\times$\,10$^{\rm 7}$~M$_{\sun}$ and 
$\sim$3.6\,$\times$\,10$^{\rm 7}$~M$_{\sun}$ place them in the mass range typically found for GMAs \citep{vog88}. Their dispersion values 
and sizes (spherical radii) range from $\sim$40~km\,s$^{\rm -1}$ to $\sim$ 67~km\,s$^{\rm -1}$ and 35~pc to 56~pc, respectively. Due to 
their large velocity dispersion, the GMA candidates do not fall on the Larson size-linewidth law for molecular clouds 
\citep[e.g.,][]{lar81,sol87} but instead lie above this relation. A comparison of the CO spectra of the GMA candidates reveals a single 
spectral peak in all GMAs not associated with gas surrounding the AGN position. The spectra of some of the GMAs in the northern part of 
the shell show a double-peaked emission line (GMA~5, 8) and broad line widths probably due to blending of the two velocity components 
(GMA~6). These three complexes are closest to the nucleus of NGC~4194. The GMAs~7 and 8 are located in the northern part of the shell; 
GMA~9 to 11 are located in the southern part. The majority of these hot spots of enhanced $^{\rm 12}$CO\,2$-$1 intensity (GMA~1, 4, 7, 
9$-$11) is actually associated with the eastern dust lane that crosses the molecular gas shell. This can also be seen in NGC~1614, for 
example, where half of the GMAs identified in CO\,2$-$1 observations are associated with dust lanes.\\
\indent
The average mass of the GMA candidates in NGC~4194 is $\sim$2.2\,$\times$\,10$^{\rm 7}$~M$_{\sun}$, the line widths average at a value of 
$\sim$120~km\,s$^{\rm -1}$, and the fitted average FWHM size (spherical radius) of the GMAs is $\sim$109~pc. The GMAs in NGC~1614 
seem to lie on a different scale. They are on average more massive (6.0\,$\times$\,10$^{\rm 7}$~M$_{\sun}$), show the same line widths 
(120~km\,s$^{\rm -1}$), and are significantly larger in size (270~pc) than the GMAs in NGC~4194 \citep{koenig13}. The size estimate is 
difficult, though. Therefore, this comparison criterion should not be overestimated. Another point to keep in mind is that the 
ring-structure in NGC~4194 is only half of the size of the ring in NGC~1614. Another galaxy bearing GMAs that are comparable to the ones 
in NGC~1614 is NGC~1097 
\citep[average size $\sim$250~pc, mass 9.2\,$\times$\,10$^{\rm 7}$~M$_{\sun}$, line width 84~km\,s$^{\rm -1}$][]{hsieh11}, whereas the 
GMAs in NGC~1365 \citep[line widths between 60 and 90~km\,s$^{\rm -1}$, masses of $\sim$10$^{\rm 6}$~M$_{\sun}$,][]{saka07} seem to be 
even less massive and have smaller line widths than the ones in NGC~4194.

\begin{figure}[th]
  \centering
    \includegraphics[width=0.4\textwidth,angle=-90]{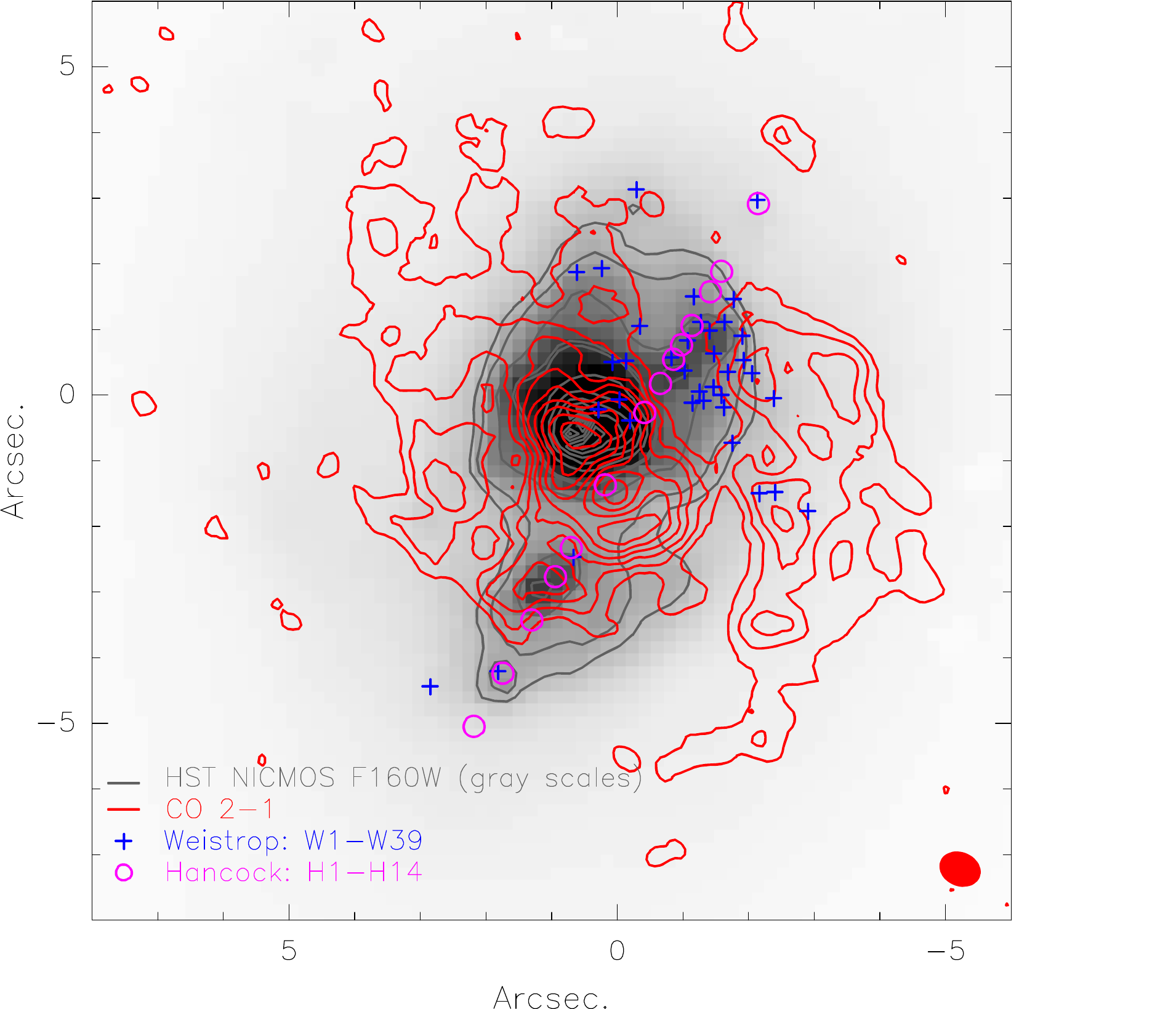}
  \caption{\footnotesize Overlay of the $^{\rm 12}$CO\,2$-$1 emission contours on top of an HST WFPC2 F606W filter image. North is up, 
   east to the left. Blue crosses indicate the positions of the star clusters that were found by \citet{wei04} in 
   the UV; magenta circles represent the star clusters identified in the visible \citep{han06}. The CO beam size is indicated in the 
   lower right corner.}
  \label{fig:hst+clusters+co21}
\end{figure}

\begin{table*}[ht]
\begin{minipage}[!h]{\textwidth}
\centering
\renewcommand{\footnoterule}{}
\caption{\small
 Properties of the identified GMA candidates.}
\label{tab:GMA_properties}
\tabcolsep0.1cm
\begin{tabular}{lccccccc}
\noalign{\smallskip}
\hline
\noalign{\smallskip}
\hline
\noalign{\smallskip}
GMA &  & RA(2000) & DEC(2000) & spherical radius\footnote{The source size was determined by fitting a Gaussian to the position of the 
respective GMA in the uv table. The values given here are the \newline \hspace*{3mm} FWHM of the fitted Gaussian deconvolved from the synthesized beam. The typical (average) error for the fit to the GMA size is about 20\%.} & $v$$_{\rm 0}$\footnote{Typical errors are on the order of 5-10~km\,s$^{\rm -1}$.} & dispersion\footnote{Typical errors are on the order of 10\%.} & mass \\
\noalign{\smallskip}
        & & [h] [m] [s] & [\degr] [\arcmin] [\arcsec] & [pc] & [km\,s$^{\rm -1}$] & [km\,s$^{\rm -1}$] & [M$_{\sun}$] \\
\noalign{\smallskip}
\hline
\noalign{\smallskip}
GMA\,1 & & 12:14:09.805 & 54:31:35.54 & 83  & 2585 & 42 & 1.0\,$\times$\,10$^{\rm 7}$\\
GMA\,2 & & 12:14:09.749 & 54:31:36.01 & 103 & 2501 & 52 & 1.5\,$\times$\,10$^{\rm 7}$\\
GMA\,3 & & 12:14:09.631 & 54:31:36.08 & 92  & 2467 & 44 & 1.2\,$\times$\,10$^{\rm 7}$\\
GMA\,4 & & 12:14:09.737 & 54:31:35.47 & 132 & 2557 & 57 & 3.4\,$\times$\,10$^{\rm 7}$\\
GMA\,5 & & 12:14:09.674 & 54:31:35.76 & 114 & 2508 & 67 & 2.8\,$\times$\,10$^{\rm 7}$\\
GMA\,6 & & 12:14:09.632 & 54:31:35.44 & 92  & 2510 & 62 & 1.6\,$\times$\,10$^{\rm 7}$\\
GMA\,7 & & 12:14:09.749 & 54:31:34.72 & 123 & 2595 & 40 & 3.1\,$\times$\,10$^{\rm 7}$\\
GMA\,8 & & 12:14:09.667 & 54:31:35.13 & 130 & 2595 & 45 & 3.5\,$\times$\,10$^{\rm 7}$\\
GMA\,9 & & 12:14:09.612 & 54:31:33.90 & 103 & 2599 & 45 & 1.9\,$\times$\,10$^{\rm 7}$\\
GMA\,10 & & 12:14:09.577 & 54:31:34.21 & 130 & 2575 & 59 & 3.1\,$\times$\,10$^{\rm 7}$\\
GMA\,11 & & 12:14:09.547 & 54:31:34.55 & 92  & 2588 & 47 & 1.3\,$\times$\,10$^{\rm 7}$\\
\noalign{\smallskip}
\hline
\end{tabular}
\end{minipage}
\end{table*}

\subsection{Circumnuclear molecular gas and star formation} \label{subsubsec:sf_molecular_gas}

\begin{figure*}[ht]
  \begin{minipage}[hbt]{0.4925\textwidth}
  \centering
    \includegraphics[width=0.82\textwidth,angle=-90]{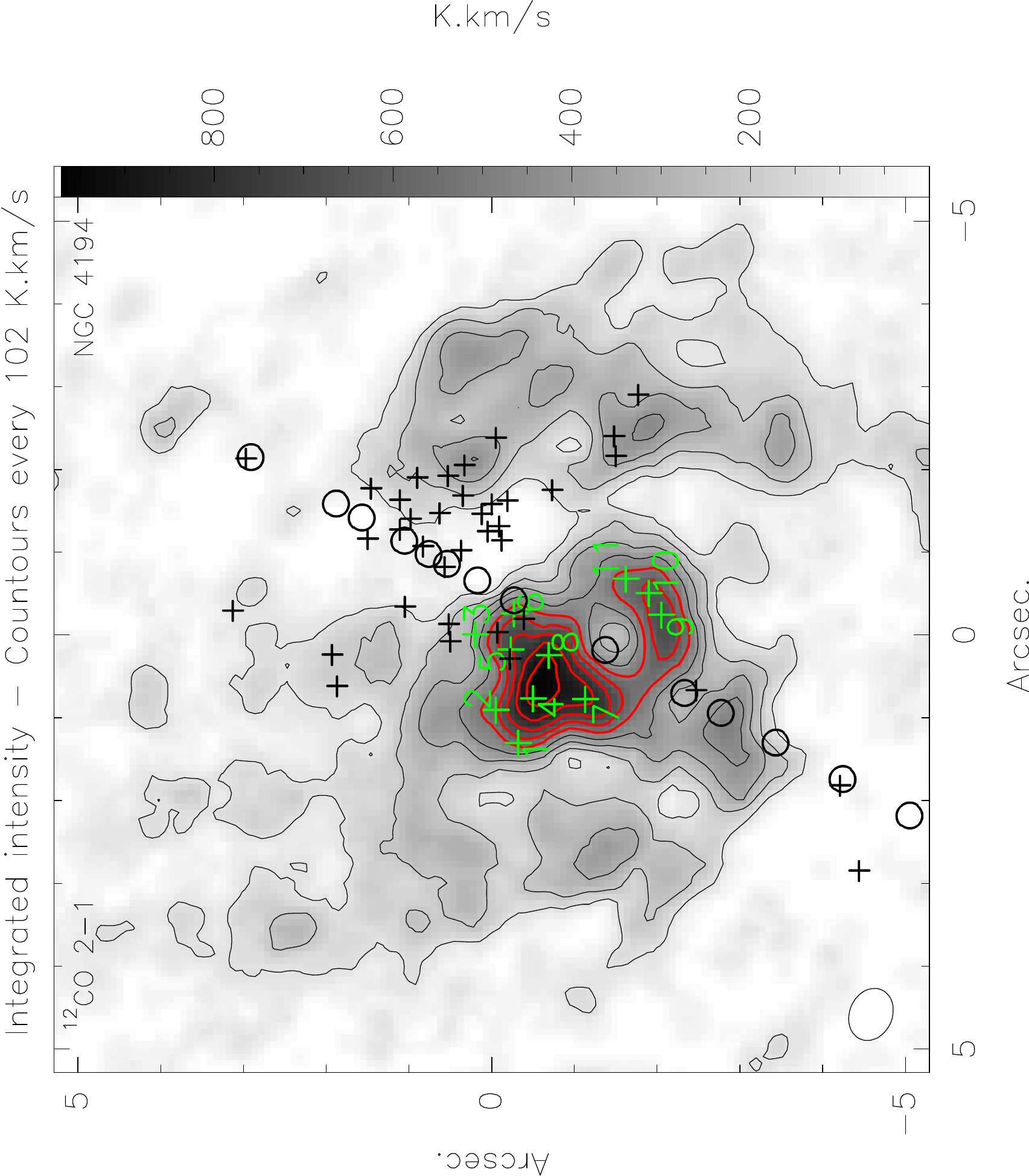}
  \end{minipage}
  \begin{minipage}[hbt]{0.4925\textwidth}
  \centering
    \includegraphics[width=0.82\textwidth,angle=-90]{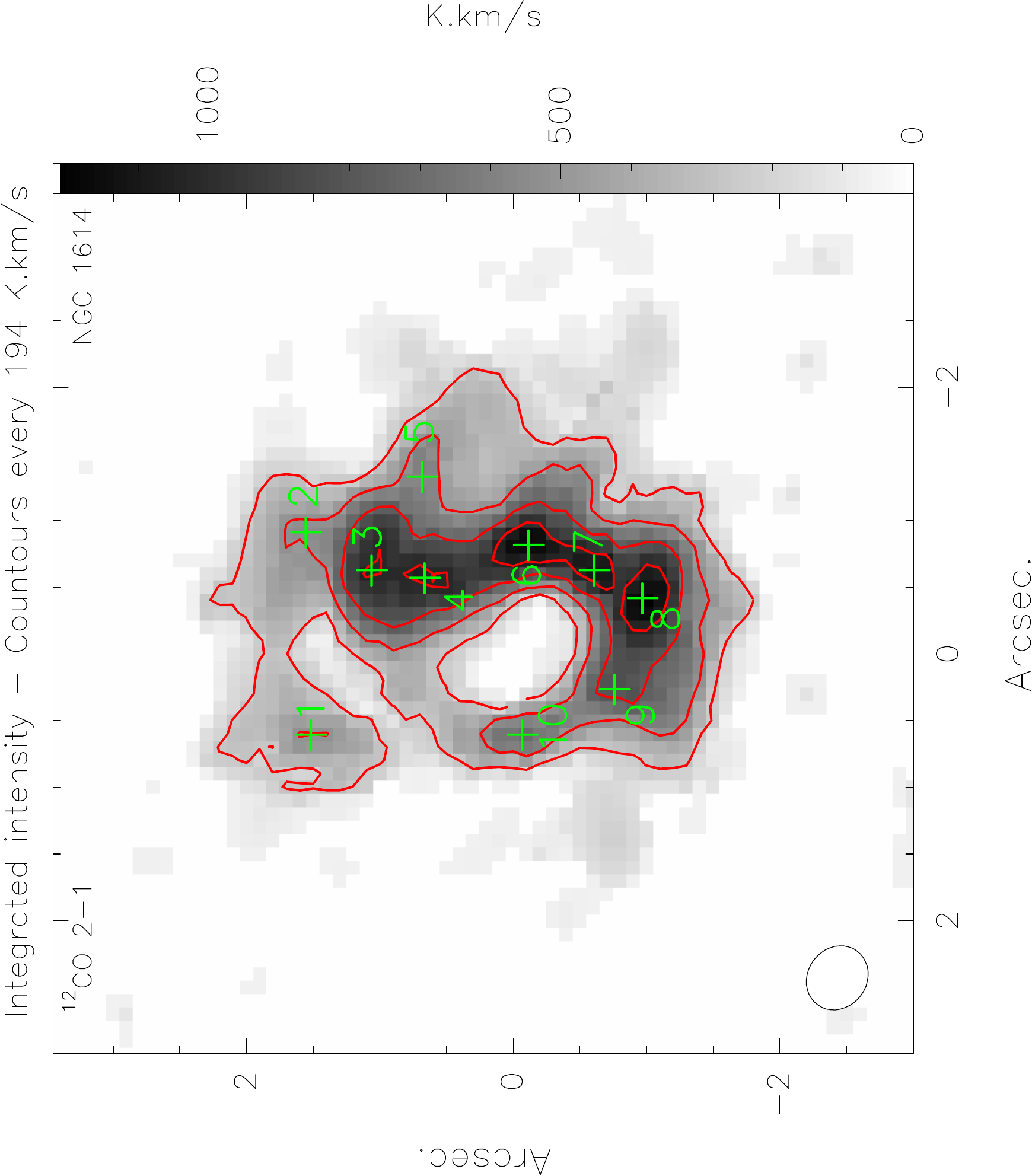}
  \end{minipage}
  \caption{\footnotesize Comparison of the nuclear molecular rings of \object{NGC~4194} (\textit{left}) and \object{NGC~1614} 
   \citep[\textit{right},][]{koenig13}. The images are scaled to represent a field of view of $\sim$2000~pc. Contours emphasized in red 
   represent the CO\,2$-$1 ring structures in the Medusa (1\arcsec\,$\sim$189~pc) and NGC~1614 (1\arcsec\,$\sim$310~pc); green crosses 
   mark the positions of the GMAs in the Medusa and NGC~1614. Black crosses represent the UV-identified star clusters \citep{wei04}, 
   and black circles indicate the star clusters identified in visible light in a 0.5\arcsec\ wide slit \citep{han06} in the Medusa.}
  \label{fig:comparison_co2-1_rings}
\end{figure*}

A number of different gas tracers has been observed toward NGC~4194 with different spatial resolutions. The CO\,1$-$0, radio continuum 
(at 3.5~cm, 6~cm and 20~cm), and HST observations (e.g., at wavelengths of 150~nm and 600~nm), for example, show brightness distribution 
peaks at the location of the AGN \citep[e.g.,][]{aalto00,con90,bes05,arm90,wei04}. Spectroscopic H$\alpha$ observations find the majority 
of the star formation going on in the central 8\arcsec\ \citep{wei12}. Several star forming clusters have been identified in the UV 
\citep[160~nm \& 200~nm,][]{wei04} and the VIS \citep[430~nm \& 780~nm,][]{han06}; some of them are associated with the AGN at the 
nucleus of NGC~4194 (Fig.\,\ref{fig:hst+clusters+co21}).\\
\indent
Most of the brightness peaks identified in the high resolution CO\,1$-$0 data \citep{aalto00} are situated in the main dust lane (peaks 
a, b, d, and e). Peaks a and b represent positions in the southern and northern part of the central high surface brightness density 
complex in the CO\,2$-$1 maps; peaks c and d are located in the northern and the southern part of the western arm. Two of the CO\,1$-$0 
peaks (b and c) seem to be counterparts to brightness maxima found in H$\alpha$ observations 
\citep[see Fig.\,\ref{fig:comparison_co1-0_co2-1},][]{arm90}. Peak b is furthermore associated with the peak position in the 
high-resolution 1.4~GHz continuum maps \citep[see, e.g., Fig.\,\ref{fig:overlays_highres_ring},][]{bes05}.\\
\indent
Our CO\,2$-$1 observations together with all these tracers indicate that a big part of the overall star forming activity seems to be 
on-going in the central few arcseconds around the AGN. However, to gain more insight into the locations and the properties of the star forming regions in the center of NGC~4194, further observations are needed. For example, high resolution Br$\gamma$ narrow-band images 
would give valuable insight into the positions of regions of massive star formation; HCN/HCO$^{\rm +}$ observations would help to 
distinguish high density regions ($n$\,$\gtrsim$\,10$^{\rm 4}$~cm$^{\rm -3}$).\\
\indent
The search for star clusters has yielded detections in the visible and in the UV wavelength regimes (see 
Figs.\,\ref{fig:hst+clusters+co21} and \ref{fig:comparison_co2-1_rings}). \cite{wei04} used HST imaging to look for star clusters 
in the UV. They identified 39 young stellar clusters between the ages of 5 and 15~Myr distributed over the the optical body of NGC~4194. 
\citet{han06} extended the properties of star clusters in this galaxy by identifying 14 clusters in a slit positioned along a position 
angle of 147\degr\ using visible and UV light. Some of the clusters from the two samples seem to coincide within the placement of the 
slit, but there are also clusters at slit positions in the visible light that seem to have no UV counterparts. The majority of the UV and 
VIS identified star clusters does not seem to be closely associated with the molecular gas. Indeed, most of these clusters are positioned 
in a void of molecular gas between the eastern and western parts in the CO\,2$-$1 emission. Most of the UV-identified clusters are 
located in the north and west of the CO\,2$-$1 distribution; none are associated with molecular gas in the east and north-east of the 
nucleus (Figs.\,\ref{fig:hst+clusters+co21}, \ref{fig:comparison_co2-1_rings}).\\
\indent
A comparison to the high resolution CO\,2$-$1 shows that the GMAs closest to the nucleus (GMA~3, 5 and 6) seem to be associated with some 
of the \citet{wei04} UV clusters (Fig.\,\ref{fig:comparison_co2-1_rings}). One of the clusters identified by \citet{han06} is also 
located in that region. The southern part of the CO shell, on the other hand, harbors no star clusters, although we identified three GMAs 
in that region. One of the VIS-identified star clusters, however, is located right at the center of the shell, where there is no CO, but 
the radio continuum has a secondary peak in this location.\\
\indent
Using the CO\,2$-$1 data, we are now able to better distinguish between the higher density gas close to the nucleus and the lower density 
gas in the surrounding environment described by the CO\,1$-$0 observations of \citet{aalto00}. Indeed, we find that the gas in the Eye 
closest to the nucleus is of higher surface brightness density than the larger scale CO emission. This reservoir of 
high-surface-brightness gas is located inside the solid body part of the rotation (Fig.\,\ref{fig:co2-1_pv}). The gas surface density 
close to the nucleus amounts to $\sim$4.1\,$\times$\,10$^{\rm 3}$~M$_{\sun}$\,pc$^{\rm -2}$, whereas we find a value of 
$\sim$1.4\,$\times$\,10$^{\rm 3}$~M$_{\sun}$\,pc$^{\rm -2}$ for the surrounding medium; hence, we find a difference of a factor of 
$\sim$3. This value would increase even more when taking the surface density of the CO\,1$-$0 
\citep[500-1000~M$_{\sun}$\,pc$^{\rm -2}$,][]{aalto00} into account instead. The difference between the obtained values from our 
CO\,2$-$1 observations and the CO\,1$-$0 data from \citet{aalto00} is that we do suffer from resolving out large parts of the lower 
brightness gas due to the high resolution of these observations.\\

\indent
The bulk of the star formation ongoing in the super star clusters that are distributed on extended scales is not associated with the 
regions of highest surface brightness density (Figs.\,\ref{fig:hst+clusters+co21}, \ref{fig:comparison_co2-1_rings}). The clusters 
identified with the highest H$\alpha$ SFR, however, are located in this high surface density region \citep{han06}. Using the CO\,2$-$1 
surface brightness density, the Kennicutt-Schmidt law predicts a star formation rate of $\sim$4.9~M$_{\sun}$\,yr$^{\rm -1}$, which within 
the error bars is in good agreement with the 6-7~M$_{\sun}$\,yr$^{\rm -1}$ value that \citet{aalto00} estimated from the FIR but is much 
below the value of $\sim$46~M$_{\sun}$\,yr$^{\rm -1}$ that \citet{han06} derive from their H$\alpha$ observations. The star formation efficiency we obtained from the highest surface brightness density CO\,2$-$1 emission is 
$\sim$1.5\,$\times$\,10$^{\rm -8}$~yr$^{\rm -1}$. Although NGC~4194 has a brightness surface density of one magnitude lower than typical 
ULIRGs, its star formation efficiencies rival that of these extraordinary objects \citep[e.g., SFE(\object{Arp~220}) 
$\sim$2\,$\times$\,10$^{\rm -8}$,][]{fra00}. Compared to values in typical mergers, the star formation in NGC~4194 is two orders of 
magnitude more efficient \citep{row99}.\\
\indent
Although the gas surface density at the center of NGC~4194 has now increased to 
$\sim$4.1\,$\times$\,10$^{\rm 3}$~M$_{\sun}$\,pc$^{\rm -2}$, these values show that the surface density even in the densest regions in 
the center is still lower by one order of magnitude compared to the gas in the centers of ULIRGs like \object{Arp~220} or object{Mrk~231} 
that have surface densities of up to $\sim$3.3\,$\times$\,10$^{\rm 4}$~M$_{\sun}$\,pc$^{\rm -2}$ \citep{bry99}. The central high surface 
brightness gas density values in NGC~4194 are comparable to numbers for other similar sources, such as NGC~1614 
\citep[$\sim$3.1\,$\times$\,10$^{\rm 3}$~M$_{\sun}$\,pc$^{\rm -2}$,][]{koenig13} and NGC~5218 
\citep[$\sim$1.8\,$\times$\,10$^{\rm 3}$~M$_{\sun}$\,pc$^{\rm -2}$,][]{cul07}.

\subsection{Infalling gas} \label{subsubsec:infalling_gas}

1.4~GHz radio continuum observations show two compact radio components at the center of NGC~4194 \citep{bes05}. They concluded that there 
is the possibility of a weak, buried AGN that contributes to the powering of the star formation. The position that is inferred from the 
radio continuum brightness peak coincides with the velocity dispersion peak, which we find in our CO\,2$-$1 data 
(Fig.\,\ref{fig:rot_curve_residuals}a). The rotation curve fitted to the CO\,2$-$1 data cube (Fig.\,\ref{fig:rot_curve_residuals}a, 
Sect.\,\ref{subsec:kinematics}) further indicated the location of the dynamical center just $\sim$0.4\arcsec\ toward the south of the 
radio continuum and the velocity dispersion peaks. This is in positional agreement within one beam size.\\
\indent
It is possible that the nucleus is fed by molecular gas returning via the dust lane. \citet{bour05} have shown that the gas in minor 
mergers that is brought in by a disturbing galaxy companion is generally found at large radii in the merger remnant. Gas located in this 
largest scale gas reservoir is then returned to the system via tidal tails, where it often ends up forming rings, which might be 
associated with or appear as dust lanes seen edge-on. In NGC~4194, a number of dust lanes has been found. The dust lanes most probably 
playing a role in the feeding of the AGN are the two large dust lanes to the east of the nucleus and a smaller one just northeast of the 
AGN. All those dust lanes are associated with the CO emission \citep[][see also Fig.\,\ref{fig:co2-1}b]{aalto00} and might funnel the CO 
from the larger scale molecular gas reservoir, as traced by the CO\,1$-$0 emission, to the center and ultimately to the AGN. We found 
evidence for this scenario for example in the western part of the molecular gas distribution in the velocity field 
(Figs.\,\ref{fig:co2-1}c, d) and the map showing the residuals from the subtraction of the rotation curve 
(Fig.\,\ref{fig:rot_curve_residuals}b), which show typical features of streaming motions as, for example, found in \object{M~51} \citep{aalto99} and \object{M~83} \citep{ran99}.\\ 
\indent
Another mode of transport providing gas to the center of NGC~4194 might be streaming motions. We found indications for streaming motions 
in the western arm of NGC~4194. Gas could be piled up in the center of the galaxy due to orbit crowding processes of inflowing gas 
streams. In the galactic potential, the gas brought in through the streams encounters shocks and migrates to new orbits accumulating in 
these locations.

\begin{table*}[t]
\begin{minipage}[!h]{\textwidth}
\centering
\renewcommand{\footnoterule}{}
\caption{\small
 Overview of the parameters of galaxies compared to NGC~4194.}
\label{tab:vgl_gal}
\tabcolsep0.1cm
\begin{tabular}{lccccccccccc}
\noalign{\smallskip}
\hline
\noalign{\smallskip}
\hline
\noalign{\smallskip}
 Galaxy & & Merger/ & \multicolumn{3}{c}{presence of} & \multicolumn{2}{c}{ring key properties} & \multicolumn{3}{c}{shell key properties} & Reference\footnote{References: 1): 
this work, 2): \citet{bec14}, 3) J\"utte et al. (in prep.), 4): \citet{koenig13}, 5): \citet{ols07}, \newline \hspace*{2.8mm} 6): \citet{gal10}, 7): \citet{spa09}, 8): \citet{kri05}, 9): \citet{man08}, 10): \citet{jue10}, \newline \hspace*{2.5mm} 11): \citet{weis99}, 12): \citet{nak87}, 13): \citet{saka06}} \\
\noalign{\smallskip}
 & & interaction  & a bar & a ring & a shell & radius [pc] & mass [M$_{\sun}$] & radius [pc] & v$_{\rm exp}$\footnote{expansion velocity} [km/s] & mass [M$_{\sun}$]  &           \\
\noalign{\smallskip}
\hline
\noalign{\smallskip}
NGC~4194 & & + & +  & +   & + & 320 & 4.4\,$\times$\,10$^{\rm 8}$ & 180 & 55 & 2.3\,$\times$\,10$^{\rm 8}$ &  1), 2), 3) \\
NGC~1614 & & + & (+)\footnote{Its presence is still under discussion.}\saveFN\tab & +  & -- & 500 & 8.3\,$\times$\,10$^{\rm 8}$ & --  & -- & -- &  4) \\
NGC~5218 & & +\footnote{Part of an interacting galaxy pair: NGC~5218\,$\rightleftarrows$\,NGC~5216, NGC~3718\,$\rightleftarrows$\,NGC~3729, M~82\,$\rightleftarrows$\,M~81.}\saveFN\inte  & + & +  & +  & 470 & 7\,$\times$\,10$^{\rm 8}$ & 150 & 30 & 7\,$\times$\,10$^{\rm 7}$  & 5), 6) \\
NGC~3718 & & +\useFN\inte  & + & -- & -- & -- & -- & --  & -- & -- & 7), 8) \\
NGC~4441 & & + & -- & (+)\useFN\tab & + & 875 & 4.1\,$\times$10$^{\rm 8}$ & \footnote{No values given in the literature for the optical shells.}\saveFN\sfn & \useFN\sfn & \useFN\sfn & 9), 10) \\
M~82     & & +\useFN\inte & + & + & + & 200 & 3\,$\times$10$^{\rm 7}$& 65\footnote{Values are for the molecular 
``superbubble''.}\saveFN\foot  & 45\useFN\foot & 8\,$\times$\,10$^{\rm 6}$ \useFN\foot & 11), 12)\\
NGC~253  & & -- & + & --  & + & -- & -- & 130 & 50 & $\sim$10$^{\rm 6}$ &  13) \\
\noalign{\smallskip}
\hline
\end{tabular}
\end{minipage}
\end{table*}


\section{Comparisons to other galaxies} \label{sec:discussion_2}

Comparisons between the features of NGC~4194 and other interacting galaxies with similar types of structure can help in the 
interpretation of our data. In this section, we consider how the properties of NGC~4194 relate to those observed in a comparison sample 
of interacting galaxies and galaxies with shells/bubbles.

\subsection{Comparison to other minor axis dust lane minor mergers} \label{subsubsec:comparison_minor_mergers}

One galaxy sharing a number of properties with NGC~4194 is the S+s minor merger \object{NGC~1614}. Both galaxies are associated with 
large reservoirs of molecular gas \citep[e.g.,][]{aalto00,ols10,koenig13}. Whereas the star formation in NGC~1614 is mostly 
associated with molecular gas (CO) and other tracers like Pa$\alpha$, radio continuum, etc. \citep{koenig13}, most of the star formation 
in NGC~4194 is taking place in star clusters \citep{wei04,han06}, which are not associated with the bulk of the $^{\rm 12}$CO emission, 
but instead, mostly with lower surface brightness gas to the west of the mergers nucleus.\\
\indent
Looking into the molecular gas content in these two galaxies in more detail, NGC~4194 and NGC~1614 share even more properties. In both 
galaxies, a large reservoir of CO\,1$-$0 emission is associated with the minor axis dust lanes crossing the mergers main body. This 
association between molecular gas and dust lanes led to suggestions that the dust lanes are playing a key role in the transport of the 
molecular gas into the centers of galaxies of this type (minor mergers). Both galaxy centers harbor ring-like molecular structures that 
are clearly identified in high-resolution $^{\rm 12}$CO\,2$-$1 observations \citep[][this work]{koenig13} that might be connected to the 
larger scale CO\,1$-$0 reservoirs via the respective dust lanes. The ring-structures are, however, different in size and mass. The ring 
in NGC~1614 is twice as large and twice as massive \citep{koenig13} than the shell complex in NGC~4194 
(Fig.\,\ref{fig:comparison_co2-1_rings}, Table\,\ref{tab:vgl_gal}). Both ring-like structures are located off the actual nuclei of their 
galaxies: the ring in NGC~1614 is located symmetrically off-nucleus with its center located right at the galaxy's nucleus 
\citep{koenig13}, whereas the high density gas complex in NGC~4194 including the shell is located asymmetrically off nucleus. Only the 
gas to the north of the shell, containing the AGN, is located at the very nucleus of this galaxy. Both rings are closely associated with 
radio continuum emission, whereas the emission in NGC~1614 is actually associated with the ring, the radio continuum in NGC~4194 has 
a secondary peak at the center of the shell but does not coincide with most of the CO emission located in the shell. It does seem as if 
the radio continuum in NGC~4194 is avoiding the regions filled with CO in the shell. Furthermore, we have identified GMAs 
(Fig.\,\ref{fig:comparison_co2-1_rings}) in both NGC~4194 and NGC~1614 - 11 in NGC~4194 and ten in NGC~1614. The GMAs in both minor 
mergers are associated with the dust lanes. In general, the GMAs in NGC~4194 are smaller in size and mass then the ones in NGC~1614. This 
is most probably due to the AGN-shell-complex in NGC~4194 itself being much smaller than the ring in NGC~1614.\\
\indent
Other nearby minor axis dust lane galaxies at different evolutionary stages in the merger sequence share some of these properties 
as well (e.g., \object{NGC~5218}, \object{NGC~3718}, and \object{NGC~4441}). The object NGC~5218 is a merger at an earlier evolutionary 
stage with a molecular ring about 2.5 times larger and twice more massive than the ring in NGC~4194 
\citep[Table\,\ref{tab:vgl_gal}][]{cul07,ols07}. Two galaxies at a later evolutionary stage in the merger sequence are NGC~3718 and 
NGC~4441. In their morphology, NGC~3718 and NGC~4441 are very similar to NGC~4194; they have tidal tails and minor axis dust lanes 
crossing the galaxies. Closely associated with these features are molecular gas reservoirs extended out to large scales 
\citep[e.g.,][]{pot04,kri05,jue10} as in NGC~4194. Like NGC~4194, NGC~3718 hosts a large number of star clusters \citep{tri06}. Both 
mergers also harbor molecular disks/ring-like structures at their centers \citep{schw85,pot04,kri05,spa09,jue10}.\\
\indent
In a nutshell, galaxies with minor axis dust lanes involved in interactions/mergers can show extended molecular gas emission over 
scales of several kpc, and many harbor molecular gas rings at their centers. There is no obvious correlation between the size and mass 
of the observed molecular rings with the stage of the interaction/merger process the comparison galaxies represent.

\subsection{Shells in other galaxies} \label{subsubsec:comparison_shells}

A number of shells and bubbles have been discovered in external galaxies using different atomic and molecular gas tracers 
\citep[e.g., HI, OH, CO,][]{weis99,wil02,saka06,ols07,arg10}. Nearby starburst galaxies known to host molecular shells/bubbles are 
\object{M~82}, \object{NGC~253}, and \object{NGC~5218}. These bubble hosts are no mergers, but M~82 and NGC~5218 are involved in 
interactions. All three of them have a bar transporting molecular gas to the galaxies center \citep{pet00,mau96,ols07} like NGC~4194 
\citep[][J\"utte et al. in prep.]{bec14}.\\
\indent
One process, that causes the formation of shells/bubbles can be the explosion of a large number of supernovae, as suggested for the 
bubbles in M~82 \citep[e.g.,][]{wil02,arg10} and the shell in NGC~5218 \citep{ols07}. It has been determined that the energy output of up 
to several thousand type II supernovae is necessary to reach the current extents that are found for these structures. An alternative 
scenario for shell/bubble formation is that they might be caused by stellar winds coming from massive super star clusters, as suggested 
for the bubbles in NGC~253 \citep{saka06}. NGC~4194 does have a large population of star clusters, but the majority of these clusters is 
located away from the bulk of the molecular gas. In particular, we do not find a large conglomerate of star clusters at the center of the 
molecular shell in NGC~4194. This points toward supernova explosions as the driving mechanism of the shell in this galaxy.\\
\indent
An in-detail comparison between the shells/bubbles in these three galaxies with the shell in NGC~4194 (for more details see 
Table\,\ref{tab:vgl_gal}) shows that the shell in NGC~4194 is older, has a larger size, expands with a higher velocity, and is more 
massive than the shells/bubbles found in the starburst galaxies M~82, NGC~253, and NGC~5218.\\
\indent
This shows that NGC~4194 shares properties with several of the comparison galaxies - for example, being a merger, having dust lanes, 
harboring shells/bubbles and/or GMAs, having asymmetric molecular ring-like structures, and massive star clusters. However, we always 
found significant differences as well. Therefore, NGC~4194 cannot be classified as a typical barred spiral galaxy, or a typical merger of 
any flavor; it remains a very interesting yet puzzling object.


\section{Summary} \label{sec:summary}

We studied the properties of the molecular gas in the center of NGC~4194 using high angular resolution $^{\rm 12}$CO\,2$-$1 emission line 
observations.

\begin{itemize}
\item[1.] We found a ring-like structure, the Eye of the Medusa, at the center of NGC~4194. A large part of the gas in this region of 
high surface brightness is associated with the major dust lane. The structure contains molecular gas associated with the AGN and a 
molecular shell located southeast of the nucleus.
\item[2.] The event causing the formation of the molecular shell southeast of the dynamical center of NGC~4194 was most probably a 
number of supernovae explosions, which now pushes the molecular gas outward. 
\item[3.] The kinematics of the molecular gas distribution suggest that gas is transported along the dust lanes to the center of 
NGC~4194, where it replenishes the gas in the Eye. Smaller dust lanes in the very center, together with the pressure 
provided by the expanding material in the shell, then funnel this gas to the immediate surroundings of the AGN. 
\item[4.] We identified individual GMAs in the Eye, which are associated with both the shell structure and the gas surrounding the 
AGN. The GMAs are not associated with star clusters previously identified in the UV and visible wavelength regimes. 
\end{itemize}
In summary, we found that NGC~4194 is a very interesting lab to study different physical properties of the molecular gas in different 
states. It is remarkable how well-ordered the velocity field looks when comparing all the different processes going on at different 
scales of this merger. We believe we found another example of a minor merger, besides NGC~1614, where molecular gas is transported via 
the dust lanes to the very center of the galaxy. There, we found molecular gas associated with the AGN at the dynamical center and with a 
shell to the southeast of the very nucleus, which is most likely formed by explosions of supernovae. The GMAs we located in the central 
molecular structure might be the result of the formation of denser gas resulting from shocks from the supernovae explosions.\\
\indent
In spite of these results, many questions remain to be answered, such as the origin of the gas surrounding the AGN, the feeding of the 
AGN itself on even smaller scales and the origin of the GMAs, or how the kinematics of cold molecular gas affect the evolution of star 
formation and nuclear activity in minor mergers in general. A more detailed study of the molecular gas content of NGC~4194 on different 
spatial scales is needed to address these questions. New studies with instruments, such as the PdBI and the SMA, will enable us to answer 
them.

\begin{acknowledgements}
      We thank the referee for useful comments. SA thanks the Swedish Research Council (grant 621-2011-414) and the Swedish National Space 
      Board (SNSB, grant 145/11:1-3) for support. JSG thanks the College of Letters \& Science, University of Wisconsin-Madison for 
      partial support of this work. The Submillimeter Array is a joint project between the Smithsonian Astrophysical Observatory and the 
      Academia Sinica Institute of Astronomy and Astrophysics and is funded by the Smithsonian Institution and the Academia Sinica. 
      MERLIN/eMERLIN is a National Facility operated by the University of Manchester at Jodrell Bank Observatory on behalf of STFC. AIPS 
      is produced and maintained by the National Radio Astronomy Observatory, a facility of the National Science Foundation operated under 
      cooperative agreement by Associated Universities, Inc. This research has made use of the NASA/IPAC Extragalactic Database (NED) 
      which is operated by the Jet Propulsion Laboratory, California Institute of Technology, under contract with the National Aeronautics 
      and Space Administration.
\end{acknowledgements}

\bibliographystyle{aa}
\bibliography{ngc4194}

\end{document}